\documentclass[amsmath,10pt,twocolumn,superscriptaddress,footinbib,pre]{revtex4-1}
\usepackage{amsmath,amsfonts,bm}
\usepackage{graphicx}
\usepackage{subfigure}
\usepackage{float}
\usepackage{siunitx}
\usepackage{uri}

\usepackage[dvipsnames,usenames]{xcolor} 

\usepackage{color}
\usepackage[colorlinks=true,allcolors=blue]{hyperref}
\usepackage{tikz}

\begin{document}
\title{Current inversion in a periodically driven two-dimensional Brownian ratchet}
\author{Nils E.~Strand}
\author{Rueih-Sheng Fu}
\author{Todd R.~Gingrich}
\email{todd.gingrich@northwestern.edu}
\affiliation{Department of Chemistry, Northwestern University, 2145 Sheridan Road, Evanston, Illinois 60208, USA}

\begin{abstract}
It is well known that Brownian ratchets can exhibit current reversals, wherein the sign of the current switches as a function of the driving frequency.
We introduce a spatial discretization of such a two-dimensional Brownian ratchet to enable spectral methods that efficiently compute those currents.
These discrete-space models provide a convenient way to study the Markovian dynamics conditioned upon generating particular values of the currents.
By studying such conditioned processes, we demonstrate that low-frequency negative values of current arise from typical events and high-frequency positive values of current arises from rare events.
We demonstrate how these observations can inform the sculpting of time-dependent potential landscapes with a specific frequency response.
\end{abstract}
\maketitle

\section{Introduction}
Brownian ratchets, or stochastic pumps, are spatially periodic, nonequilibrium systems that harness stochastic fluctuations to generate currents and perform useful work~\cite{reimann2002brownian, astumian2002brownian, hanggi2009artificial}.
Such ratchets have served as models for cellular motion~\cite{peskin1993cellular, mahmud2009directing}, motor proteins such as myosin, dynein, and kinesin~\cite{vale1990protein, cordova1992dynamics, astumian1994fluctuation, astumian1996mechanochemical, astumian1999chemically, huxley2000role, chowdhury2005physics, ritort2006single, vondelius2011walking, romanczuk2012active, bressloff2013stochastic, hoffmann2016molecular}, DNA-bound proteins~\cite{cocco2014stochastic}, molecular pumps~\cite{astumian1998fluctuation, astumian2001towards, siwy2002fabrication}, and artificial molecular motors~\cite{hernandez2004reversible, kottas2005artificial, chatterjee2006beyond, kay2007synthetic, serreli2007molecular, ruckner2007chemically, erbas2015artificial}.
That random noise can be rectified into work away from equilibrium---the \textit{ratchet effect}~\cite{reimann2002brownian}---is remarkable in light of equilibrium results to the contrary: Brillouin's paradox~\cite{brillouin1950can, sokolov1998energetics}, the Smoluchowski-Feynman ratchet~\cite{smoluchowski1927experimentell, feynman2011feynman}, and Parrondo's games~\cite{harmer2001brownian} in the respective contexts of circuits, mechanics, and game theory.

Theoretical analyses of one-dimensional, single-particle transport have been immensely productive at revealing the essential ratcheting mechanisms that enable rectification~\cite{reimann2002brownian, astumian1994fluctuation, doering1994nonequilibrium, astumian1996mechanochemical, astumian1996adiabatic, bier1996biased, tarlie1998optimal, Rozenbaum2008, Rozenbaum2014}.
One of the more nontrivial features of such ratchets is that the current can be remarkably sensitive to specific tunable parameters, leading to current reversals: past certain critical values, it is possible for the current to switch sign.
This type of phenomenon has largely appeared in the contexts of deterministic inertial ratchets~\cite{mateos2002current, mateos2003current, vincent2010current}, superconducting vortex ratchets~\cite{villegas2003superconducting, lu2007reversible}, and even quantum ratchets~\cite{reimann1997quantum, reimann1998quantum, lau2016identification}.
Brownian ratchets have also revealed current reversals in response to variations of parameters such as the driving frequency~\cite{bartussek1994periodically, elston1996numerical, reimann2001introduction, wickenbrock2011current}, the noise~\cite{millonas1994transport, doering1994nonequilibrium, bier1996biased, kula1998brownian, zeng2010current}, the shape of the energy landscape~\cite{reimann1996brownian}, and the particle-particle interaction strength~\cite{liebchen2012interaction}.
Sometimes even multiple inversions have been reported~\cite{kostur2001multiple, tammelo2002three, cubero2010current, lau2016identification}.

The goal of this paper is to elucidate the origin of a driving frequency-induced current reversal via the classical stochastic dynamics of a single particle.
That this phenomenon occurs can be traced back to the profoundly nonequilibrium nature of the dynamics, which relaxes into a time-periodic steady state rather than a thermal equilibrium.
The current-generating cycles of this steady state require escape events that help the system overcome energetic barriers~\cite{astumian1994fluctuation, astumian1996mechanochemical, bier1996biased}, and the escape mechanism that kinetically dominates depends on the driving frequency~\cite{astumian1996adiabatic}.

Though analytical studies of one-dimensional ratchets have illuminated the basic theoretical picture, escape over energy barriers can depend on dimensionality.
Even when current is measured along a single dimension, the particles themselves can often move in multiple dimensions~\cite{derenyi1998ac, eichhorn2003absolute, dasilva2008reversible, mcdermott2016collective}.
Inspired by their experimental measurements of currents in electron ratchets~\cite{kedem2017drive,kedem2017light,lau2020electron}, Kedem \textit{et al.} have begun addressing the importance of dimensionality via classical simulations involving driven, damped Langevin dynamics on a two-dimensional transport layer~\cite{kedem2017mechanisms}.
These studies highlighted that adding a second degree of freedom allows for a symmetry-breaking mechanism inaccessible to one-dimensional ratchets, motivating further investigation beyond one-dimensional toy models.
Using an ensemble of simulated Langevin trajectories to analyze behavior very close to the current reversal, however, can be numerically challenging.
Because the magnitude of the current is necessarily small near an inversion, detecting signal from noise becomes particularly costly.

In this work, we set out to develop a two-dimensional lattice model that would bypass continuous-space, discrete-time Langevin simulation.
This lattice model, which reduces to overdamped Langevin dynamics in the continuum limit, replaces trajectory simulations with spectral calculations, obviating the noise and expense of sampling.
We use this model to probe the dependence of the ratchet current on the driving frequency, allowing us to identify characteristic trajectories for the high- and low-frequency regimes as well as the crossover between classes of trajectories at the current-reversal frequency.
Further, we provide a qualitative explanation for trends in the ratchet current as the driving frequency increases and identify the impact of the potential on various aspects of such trends.
We envision that such numerical calculations could help sculpt spatiotemporal driving protocols to generate ratchets with a targeted dynamical response.

\section{Methods}
\label{sec:methods}

\subsection{Experimental system and model}
\begin{figure}[htb]
    \centering
    \begin{tikzpicture}
        \node at (-0.7,0) {\includegraphics[width=0.28\textwidth]{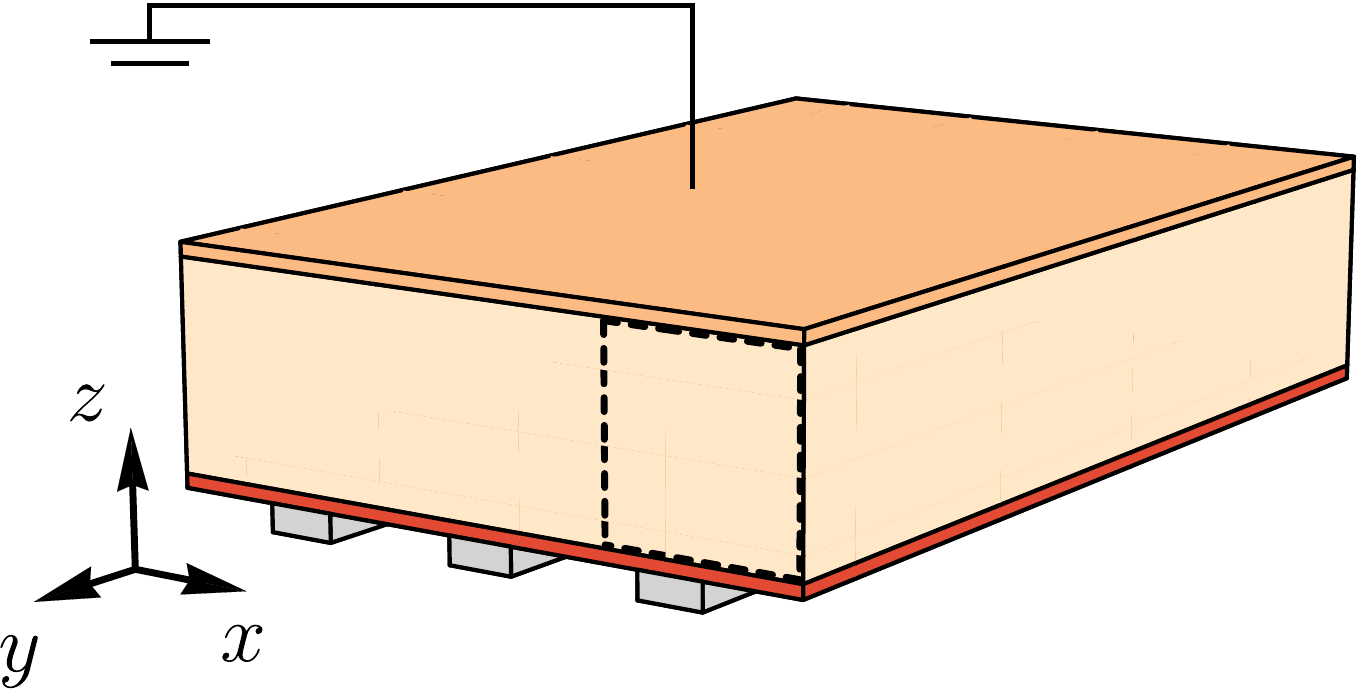}};
        \node at (3.6,0) {\includegraphics[width=0.18\textwidth]{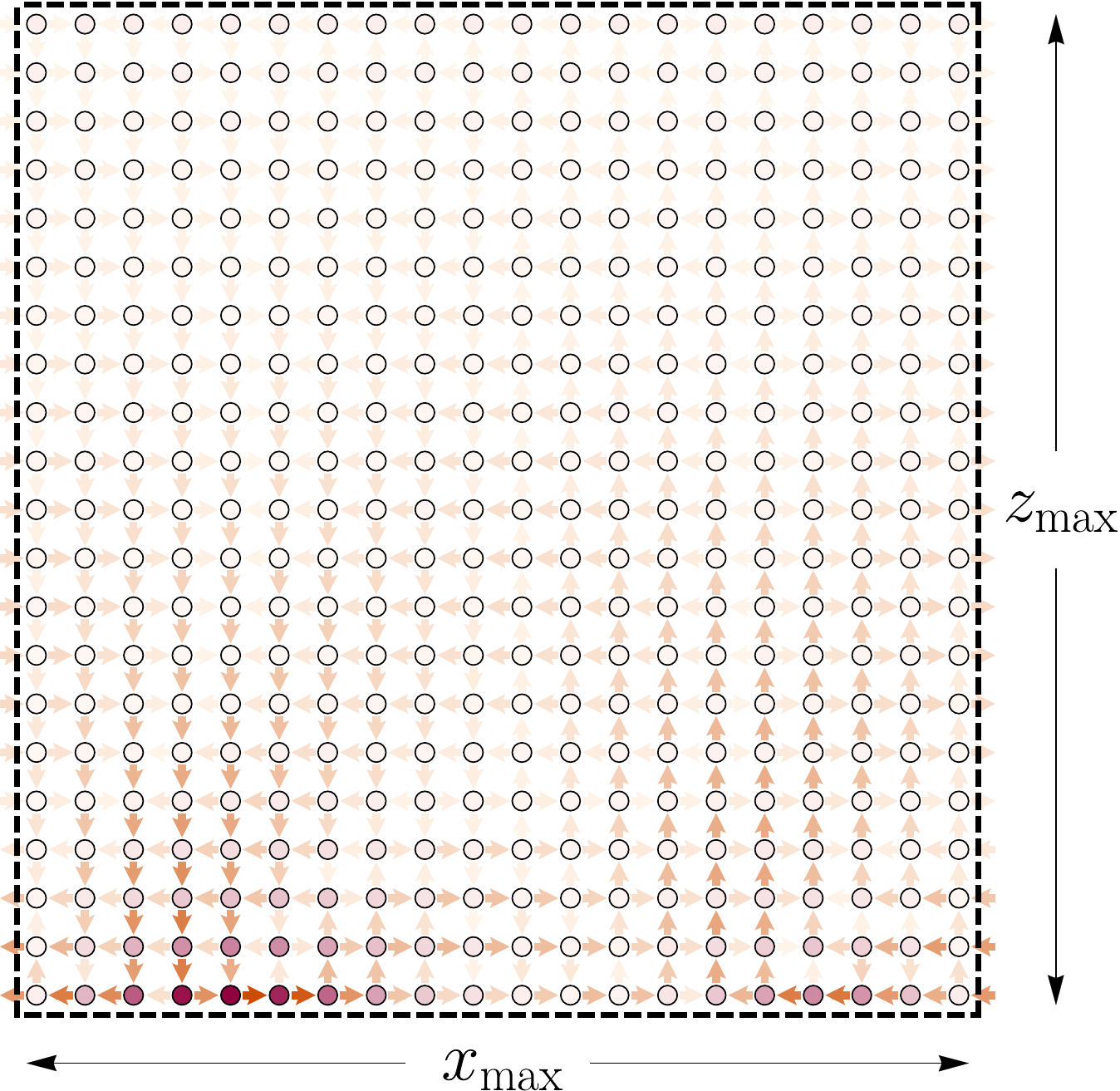}};
        \node at (-3,1.4) {(a)};
        \node at (1.7,1.4) {(b)};
    \end{tikzpicture}
    \caption{Schematic of the ratchet.
      (a) In accordance with Eqs.~\eqref{topboundary} and~\eqref{bottomboundary}, the top peach tile is grounded, while the bottom red tile is subject to a time-dependent electrostatic potential.
      Charges are mobile within the beige transport layer. 
      When averaged over a temporal period, the charge can be transported along the periodically replicated \(x\) direction.
      Because motion in the \(y\) direction cancels out on average, it suffices to study motion projected into the two-dimensional cross section outlined by the dashed black rectangle.
      (b) Example of the spatial discretization scheme used to analyze dynamics in that two-dimensional cross section, showing probability densities (shaded circles) and currents (arrows) on a 20-by-21 grid.}
    \label{model}
\end{figure}

Following Kedem~\textit{et al.}~\cite{kedem2017drive,kedem2017light}, we study classical transport in a two-dimensional cross section of a three-dimensional device, as depicted in Fig.~\ref{model}.
The device is constructed so that the electrostatic potential along the top and bottom of the device can be controlled in both space and time.
In particular, the top surface is grounded and the potential on the bottom surface tuned using metal finger electrodes beneath the transport layer.
These electrodes are periodically spaced along the \(x\) direction and run parallel to the \(y\) axis.
We assume an infinitely long device in the \(y\) direction, thereby allowing us to neglect edge effects.
Since the translational symmetry along \(y\) renders irrelevant any diffusive motion parallel to the length of the electrodes, motion can be projected solely onto the \(xz\) plane.
By charging and discharging the electrodes, a spatiotemporal electrostatic potential \(U(x,{z = 0},t)\) can be imposed along the bottom surface of the plane.
We model the electrode array by considering a single electrode and applying periodic boundary conditions along the \(x\) direction, the direction of electronic transport.

\begin{figure*}[htb]
    \centering
    \begin{tikzpicture}
        \node at (0,6.4) {\includegraphics[width=0.45\textwidth]{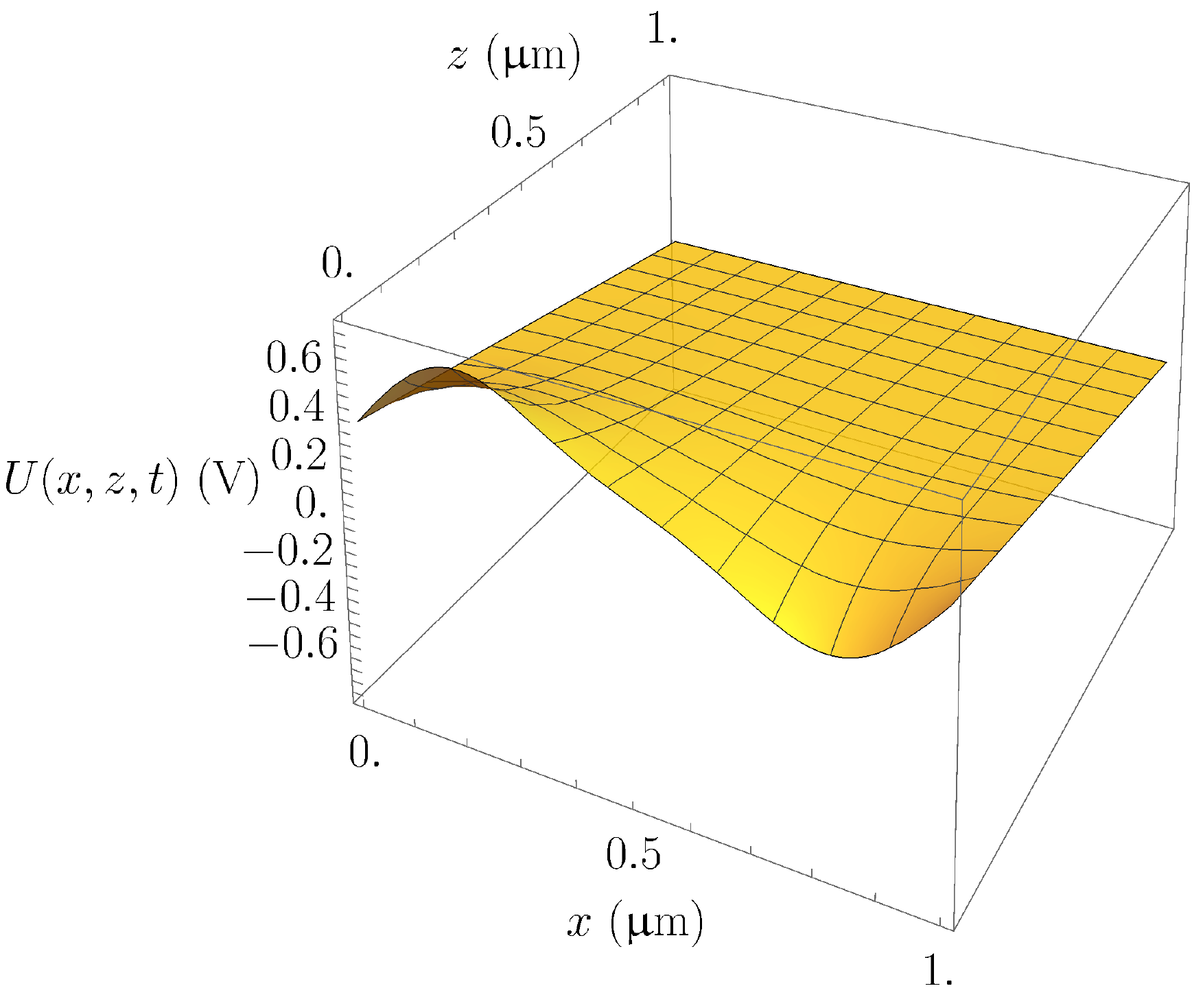}};
        \node at (8.5,6.4) {\includegraphics[width=0.45\textwidth]{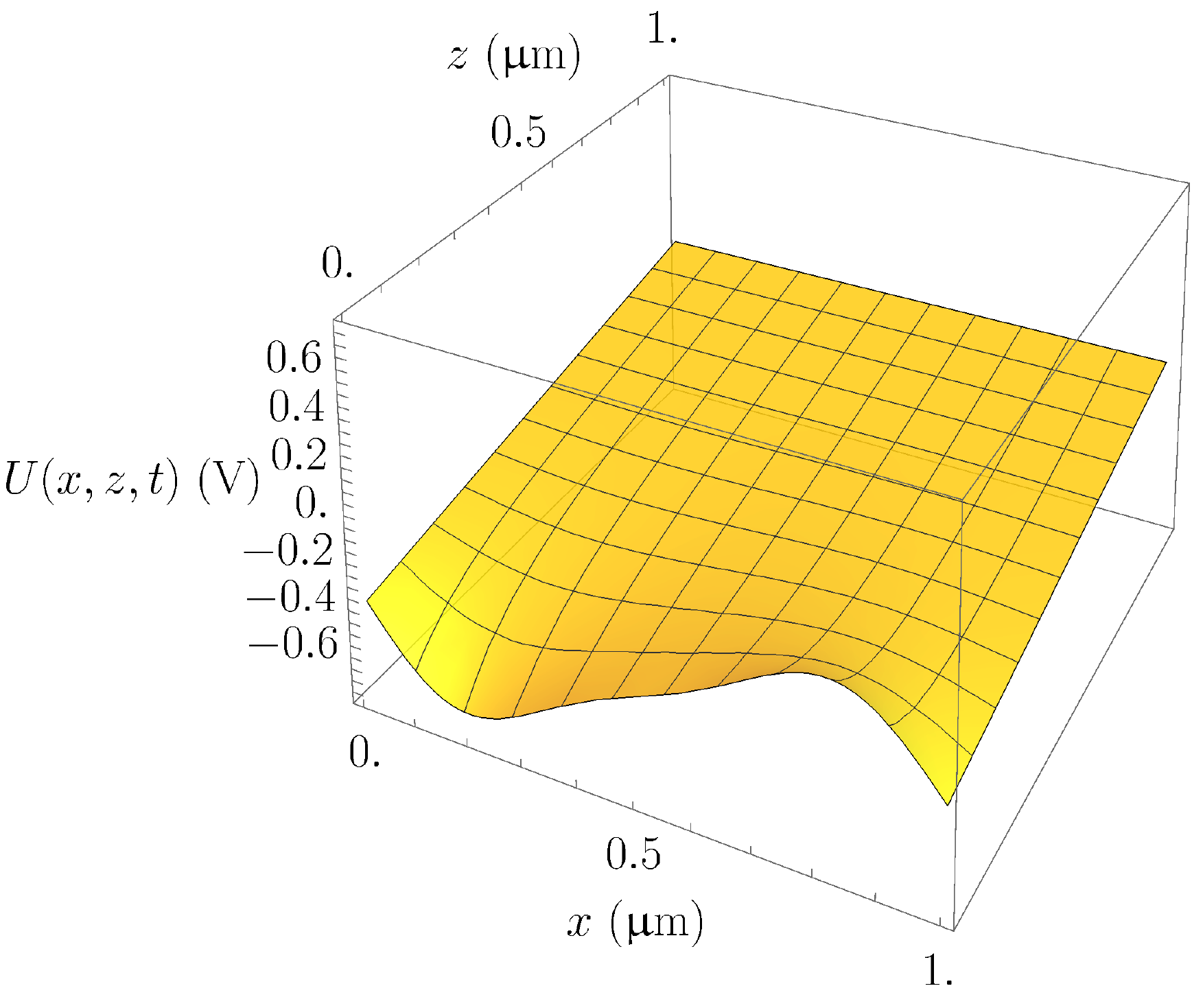}};
        \node at (-2,8.8) {(a)};
        \node at (6.5,8.8) {(b)};
    \end{tikzpicture}
    \caption{The electrostatic potential landscape \(U\) flips between two landscapes: (a) one at times \(0\leq t < \tau/2\) and (b) the other at times \(\tau/2 \leq t < \tau\).}
    \label{twopots}
\end{figure*}

The electrostatic potential throughout the transport layer follows by solving Laplace's equation \(\nabla^2 U = 0\) subject to periodic boundary conditions in the \(x\) direction and the boundary conditions
\begin{equation}
    U(x,{z=z_\text{max}},t) = 0 \label{topboundary}
\end{equation}
and
\begin{equation}
    U(x,{z=0},t) = X(x)T(t) \label{bottomboundary}
\end{equation}
along the top (\({z=z_\text{max}}\)) and bottom (\({z=0}\)) surfaces, respectively. 
In Eq.~\eqref{bottomboundary}, \(T(t)\) is a function periodic in time with period \(\tau\), while \(X(x)\) is periodic in space.
We specialize to the case that \(T(t)\) is an odd, periodic square wave with amplitude \(V_{\rm max}\),
\begin{equation}
T(t) = \begin{cases}
    V_{\rm max}, & 0 \leq t < \frac{\tau}{2}\\
    -V_{\rm max}, & \frac{\tau}{2} \leq t < \tau.
    \end{cases}
\end{equation}
Following the setup of \cite{kedem2017mechanisms}, we consider the spatial potential
\begin{equation}
    X(x) = \frac{a_1+a_2}{2}+\frac{a_1}{2}\sin\left(\frac{2\pi x}{x_\text{max}}\right)+\frac{a_2}{2}\sin\left(\frac{4\pi x}{x_\text{max}}\right),
    \label{eq:Xx}
\end{equation}
where \(x_\text{max}\) is the spacing between the periodic metal electrodes and \(a_1\) and \(a_2\) parametrize the applied potential. 
For ease of comparison, we use the same numerical parameters as \cite{kedem2017mechanisms}: \(a_1 = 1\), \(a_2 = 0.25\), \(x_\text{max} = z_\text{max} = 1\)~\si{\um}, and \(V_{\rm max} = 0.6\)~\si{\V}.
Solving the boundary value problem by separation of variables yields the exact potential,
\begin{align}
    \label{eq:exactsoln}
    \nonumber U(x,z,t) &= T(t)\times \left[\frac{a_1+a_2}{2}\left(1-\frac{z}{z_\text{max}}\right)\right.\\
    &\hspace{-7ex}+\left.\sum_{n=1}^2\frac{a_n}{2}\frac{\sin(k_nx)\sinh(k_n(z_\text{max}-z))}{\sinh(k_nz_\text{max})}\right],
\end{align}
where \(k_n=2n\pi/x_\text{max}\).
Given the temporal square-wave drive, the potential experienced by a carrier thus periodically switches between the two landscapes plotted in Fig.~\ref{twopots}.

As we will see, this potential supports a nonzero current, even though intuition from one-dimensional ratchets might lead one to think currents would vanish due to symmetry.
Indeed, if transport were constrained to a single dimension and driven by the potential we apply along the \(z = 0\) boundary condition, \(U(x,t) = X(x)T(t)\), it is well established that no current would be generated~\cite{ajdari1994rectified, tarlie1998optimal, reimann2002brownian, rozenbaum2019symmetry}.
In this case, whatever current moves along the positive-\(x\) direction at time \(t\) would be exactly counteracted by current moving in the negative-\(x\) direction at time \(\tau - t\), where \(\tau\) is the period of \(T(t)\).
However, when the carriers are allowed to move along the \(z\) dimension as well, the linear tilt \((1-z/z_\text{max})(a_1 + a_2) / 2\) of Eq.~\eqref{eq:exactsoln} now induces cycling along the second dimension \(z\).
This cycling in \(z\) ensures that current along the \(x\) direction generated at an early time \(t\) in the driving period is not exactly canceled out by the countervailing current at time \(\tau - t\)~\cite{kedem2017mechanisms}.
Consequently, although current is symmetry forbidden in one dimension, it is allowed in the two-dimensional transport layer.

\subsection{Discretization in space and time}
Overdamped dynamics on the potential landscape can be described both in Langevin form, 
\begin{equation}
    \dot{\mathbf{x}} = \mu\mathbf{f}(\mathbf{x},t) + \boldsymbol{\xi}(t), \; \langle\xi_i(t)\xi_j(t')\rangle = 2D\delta_{ij}\delta(t-t'),
\end{equation}
and in Fokker-Planck form,
\begin{equation}
    \frac{\partial\rho(\mathbf{x},t)}{\partial t} = -\mu\nabla\cdot(\mathbf{f}(\mathbf{x},t)\rho(\mathbf{x},t))+D\nabla^2\rho(\mathbf{x},t),
\end{equation}
where \(\mathbf{x} = [x \; z]\), \(\rho\) is the probability density, \(\mu\) the mobility, \(D\) the diffusion constant, \(\boldsymbol{\xi}(t)\) a random Gaussian noise, and \(\mathbf{f}(\mathbf{x},t)\) a deterministic force, given in our case by \(-\nabla U\).
The Langevin form is naturally discretized in time as
\begin{equation}
    \mathbf{x}_{i+1} = \mathbf{x}_i + \mu\Delta t\,\mathbf{f}(\mathbf{x}_i) + \sqrt{2D\Delta t}\,\boldsymbol{\eta}_i, \label{discretizedLangevin}
\end{equation}
where \(\Delta t\) is a discrete time step, \(\mathbf{x}_i \equiv \mathbf{x}(i\Delta t)\), and \(\boldsymbol{\eta}_i \sim N(0,1)\) is a random vector drawn from a unit normal distribution.
Numerical propagation of the overdamped Langevin equation given a specific noise process \(\{\boldsymbol{\eta}\}\) generates a single trajectory.
Sampling \(N\) independent trajectories produces an estimate of the horizontal current \(\bar{\jmath}_x\) with uncertainty in the estimated value decaying, in accordance with the central limit theorem, as \(N^{-1/2}\).
This approach formed the basis for previous numerical studies of this two-dimensional ratchet \cite{kedem2017mechanisms,kedem2019cooperative}.

Alternatively, one may coarse grain in space and model the dynamics as a nearest-neighbor Markov jump process on a grid of discrete spatial configurations.
Such a process is fully characterized by the master equation
\begin{equation}
    \frac{\partial\mathbf{p}}{\partial t} = \mathsf{W}\mathbf{p},
    \label{eq:master}
\end{equation}
where \(\mathsf{W}\) is the rate matrix whose elements are the transition rates between each pair of lattice sites, and \(\mathbf{p}\) is a vector whose \(i^\text{th}\) component gives the probability that the \(i^\text{th}\) lattice site is occupied.
The steady-state solution of the master equation is denoted \(\boldsymbol\pi\), which satisfies \(\mathsf W\boldsymbol\pi=\mathbf 0\).
In two-dimensional space, the rate matrix \(\mathsf W\) may be constructed to ensure that the continuum Fokker-Planck description is obtained in the limit that the grid spacing \(h\) tends to \(0\)~\cite{gardiner2009stochastic, gingrich2017inferring, bou2018continuous}.
In particular, we require that the first two moments of the rate matrix have the correct drift and diffusion:
\begin{align}
    \mu\mathbf{f}(\mathbf{x},t) &= \sum_{\mathbf{x}'}(\mathbf{x}'-\mathbf{x})\mathsf{W}_{\mathbf{x}\to\mathbf{x}'}(t;h),\\
    2D\mathsf{I} &= \sum_{\mathbf{x}'}(\mathbf{x}'-\mathbf{x})\otimes(\mathbf{x}'-\mathbf{x})\mathsf{W}_{\mathbf{x}\to\mathbf{x}'}(t;h),
\end{align}
where \(\mathsf{I}\) denotes the identity matrix and \(\mathsf{W}_{\mathbf{x}\to\mathbf{x}'}(t;h)\) is the time-dependent transition rate from \(\mathbf{x}\) to \(\mathbf{x}'\) as parametrized by \(h\). These two tensor equations decouple into a set of four scalar equations, from which the transition rates right, left, up, and down follow as
\begin{align}
\nonumber
    \mathsf{W}_{\rightarrow} &= +\frac{\mu f_1(\mathbf{x},t)}{2h}+\frac{D}{h^2},\\
\nonumber
    \mathsf{W}_{\leftarrow} &= -\frac{\mu f_1(\mathbf{x},t)}{2h}+\frac{D}{h^2},\\
\nonumber
    \mathsf{W}_{\uparrow} &= +\frac{\mu f_2(\mathbf{x},t)}{2h}+\frac{D}{h^2},\\
    \mathsf{W}_{\downarrow} &= -\frac{\mu f_2(\mathbf{x},t)}{2h}+\frac{D}{h^2}
\label{eq:rates}
\end{align}
with \(\mathbf{f} = [f_1\;f_2]\)~\cite{gingrich2017inferring}.
The same grid spacing \(h\) is used along both the \(x\) and \(z\) directions, though that choice may easily be relaxed.
In accordance with \cite{kedem2017light}, we choose \(\mu\) to be 0.005~\si{\cm\squared\per\V\per\s}, from which the diffusion constant of 12.64~\si{\um\squared\per\ms} can be obtained via the Einstein relation \(D = \mu k_\text{B}T / |q|\), where \(T\) is the system temperature, \(k_\text{B}\) the Boltzmann constant, and \(q\) the electron charge.
Unless otherwise specified, single particles are allowed to hop on a 100-by-101 lattice with a grid spacing \(h\) of 10~\si{\nm}~\footnote{The number of lattice sites differs in the \(x\) and \(z\) directions because the \(x\) direction is periodically replicated, while the \(z\) direction is not.}.

To ensure convergence of either the continuous-space Langevin approach or discrete-space jump process, the time step \(\Delta t\) or the grid spacing \(h\) must be made sufficiently small.
Appendix~\ref{convergence} addresses how fine of discretization is required as a function of both \(D\) and \(V_{\rm max}\).
There, it is shown that as \(V_{\rm max}\) increases, discretizing in time becomes advantageous, but when \(D\) increases, it becomes preferable to discretize in space.
Crucially, when \(V_{\rm max}\) is sufficiently small that spatial calculations are practical, currents can be computed using spectral methods that do not suffer from the noise of trajectory sampling.
In the next section, we describe those spectral methods in detail.

\subsection{Currents from spectral calculations}
\label{sec:spectral}
The starting point for the spectral calculations is the time-dependent rate matrix \(\mathsf{W}(t)\), with rates given by Eq.~\eqref{eq:rates}.
\(\mathsf{W}(t)\) is a sparse \(N\)-by-\(N\) matrix.
Our temporal square wave driving \(T(t)\) results in periodic toggling between 
one set of rates, \(\mathsf{W}_1\), and another, \(\mathsf{W}_2\), each for duration \(\tau/2\); that is,
\begin{equation}
    \mathsf{W}(t) = \begin{cases} \mathsf{W}_1, & 0 \leq t < \tau/2,\\ \mathsf{W}_2, & \tau/2 \leq t < \tau. \end{cases}
    \label{ratemat}
\end{equation}
General forms of \(\mathsf{W}(t)\) for arbitrary time-dependent potentials could similarly be developed as a limit of piecewise-constant rate matrices.

The temporal evolution of the steady-state state vector \(\boldsymbol{\pi}(t)\), whose \(i^\text{th}\) component gives the probability that the \(i^\text{th}\) lattice site is occupied at time \(t\), is readily obtained from \(\mathsf{W}(t)\).
Starting from \(t = 0\), the state vector after one period is
\begin{equation}
\boldsymbol{\pi}(\tau) = \mathsf{T} \boldsymbol{\pi}(0) \equiv e^{\tau \mathsf{W}_2 / 2} e^{\tau \mathsf{W}_1 / 2} \boldsymbol{\pi}(0),
\label{eq:periodpropagation}
\end{equation}
where \(\textsf{T}\) is the full-period transition matrix for the system.
After \(n\) periods, \(\boldsymbol{\pi}(n \tau)\) is given by the top right eigenvector of \(\mathsf{T}\).
All other eigenvectors correspond to smaller eigenvalues and hence to transient phenomena irrelevant in the steady state.
When the elements of \(\mathsf{W}_1\) and \(\mathsf{W}_2\) have sufficiently small magnitudes that the matrices may be exponentiated numerically, \(\mathsf{T}\) is readily computed and its largest eigenvalue obtained via the Arnoldi or power iteration methods.
As stated, multiplication by \(\textsf{T}\) only yields \(\boldsymbol{\pi}\) at intervals of the period \(\tau\). To obtain values of \(\boldsymbol{\pi}\) within a period, we propagate it by a fraction of \(\tau\) as follows:
\begin{equation}
\boldsymbol{\pi}(t) = \begin{cases}
e^{\mathsf{W}_1 t} \boldsymbol{\pi}(0), & 0 \leq t < \frac{\tau}{2}\\
e^{\mathsf{W}_2 \left(t - \frac{\tau}{2}\right)} e^{\mathsf{W}_1 \tau / 2} \boldsymbol{\pi}(0), & \frac{\tau}{2} \leq t < \tau.
\end{cases}
\label{eq:interiorprobs}
\end{equation}
The time-dependent steady-state current passing along an edge of the lattice from site \(k\) to neighboring site \(l\) is given simply by \(j_{lk}(t) = \pi_k(t) \mathsf{W}_{lk}(t) - \pi_l(t) \mathsf{W}_{kl}(t)\).

Suppose, however, that we do not want to resolve the temporal variations of the currents, and instead care about a period-averaged macroscopic current whose microscopic edge currents are weighted by a matrix \(\mathbf d\),
\begin{equation}
    \bar{\jmath} = \frac{1}{\tau} \int_0^\tau \text{d}t\,\sum_{kl} d_{lk} j_{lk}(t).
    \label{eq:jd}
\end{equation}
Being a current, \(\bar{\jmath}\) must switch signs upon reversal of time.
Here, \(\bar{\jmath}\) is a \emph{generalized scalar current} which averages over both time and space.
For example, \(\bar{\jmath}\) is the net particle current in the \(x\) direction when we set
\begin{equation}
    d_{lk} = \begin{cases} +1, \quad &\text{\(k\) directly to the left of \(l\),}\\ -1, \quad &\text{\(k\) directly to the right of \(l\),}\\ 0, \quad &\text{otherwise.} \end{cases}
\end{equation}
Directly computing Eq.~\eqref{eq:jd} requires integrating \(j_{lk}(t)\) over all \(t\) within a period.
One can alternatively obtain the mean and variance of that period-averaged current by computing, via spectral tools, the scaled cumulant-generating function (SCGF),
\begin{equation}
    \psi_{\bar{\jmath}}(\lambda) := \lim_{n \to \infty} \frac{1}{n} \ln \langle e^{\lambda n \bar{\jmath}}\rangle_n,
\end{equation}
where the expected value \(\langle\cdot\rangle_n\) is taken over all possible \(n\)-period trajectories.
Knowledge of \(\psi_{\bar{\jmath}}(\lambda)\) yields all cumulants of \(\bar{\jmath}\); in particular,
\begin{equation}
    \langle\bar{\jmath}\rangle = \left.\frac{\text{d}\psi_{\bar{\jmath}}}{\text{d}\lambda} \right|_{\lambda = 0} \text{  and  } \langle\delta\bar{\jmath}^2\rangle = \left.\frac{1}{n}\frac{\text{d}^2\psi_{\bar{\jmath}}}{\text{d}\lambda^2} \right|_{\lambda = 0},
\label{eq:twocumulants}
\end{equation}
where \(\delta \bar{\jmath} = \bar{\jmath} - \langle\bar{\jmath}\rangle\).
Casting the period-averaged current statistics in terms of the SCGF can be useful because \(\psi_{\bar{\jmath}}(\lambda)\) is practically computed as the maximal eigenvalue of a product of matrix exponentials~\cite{lebowitz1999gallavotti, lecomte2007thermodynamic, touchette2009large, chabane2020periodically},
\begin{equation}
\psi_{\bar{\jmath}}(\lambda) = \frac{1}{\tau} \ln \max {\rm eig} \left(e^{\mathsf{W}_2(\lambda) \tau / 2} e^{\mathsf{W}_1(\lambda) \tau / 2}\right),
\label{eq:perron}
\end{equation}
where the so-called tilted rate matrices \(\mathsf{W}_m(\lambda)\) are constructed from the original rate matrices \(\mathsf{W}_m\) as
\begin{equation}
    [\mathsf{W}_m(\lambda)]_{lk}:= [\mathsf{W}_m]_{lk} e^{\lambda d_{lk}}.
    \label{eq:tiltelements}
\end{equation}
Practically, the mean period-averaged current \(\langle\bar{\jmath}\rangle\) is computed by evaluating the maximum eigenvalue of the tilted matrix in the limit of small \(\lambda\).
Such a spectral approach efficiently enables computation of the mean current as a function of system parameters (driving field strength, diffusion constant, \textit{etc.}) without detailed attention to the mechanism of transport.
Rather than focus on the trajectories, mean currents are quickly extracted from a single eigenvalue calculation.

Though we have presented a spectral technique in the special case of square-wave potentials, the methodology generalizes naturally by approximating an arbitrary time-dependent rate matrix \(\mathsf{W}(t)\) as a collection of infinitesimal piecewise-constant rate matrices. A more compete derivation of that generalization is presented in Appendix~\ref{SCGF}.

\section{Results}
\label{sec:results}

\subsection{A subtle current reversal}
\begin{figure}[htb]
    \centering
    \includegraphics[width=0.45\textwidth]{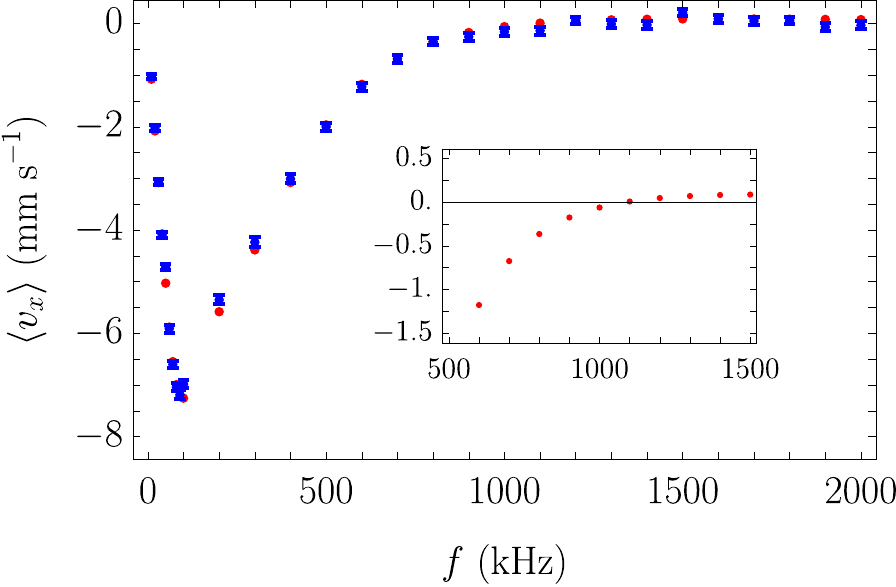}
    \caption{Current as a function of driving frequency.
      Average horizontal particle velocities computed for driving frequencies ranging from 10 to 2000~\si{\kHz}.
      Currents were calculated using both spectral methods (red) and from single-particle Langevin simulations (blue) as described in the text.
      Spectral calculations were performed on a 100-by-101 grid.
      Langevin simulations were averaged over 512 independent 10~\si{\ms} trajectories with a time step \(\Delta t\) of 35~\si{\ps}.
      Error bars represent the standard error of the mean.
      The inset reveals a subtle current reversal at a frequency of approximately \(1100~\si{\kHz}\).}
    \label{langvtilt}
\end{figure}

We employed the spectral calculations of Sec.~\ref{sec:spectral} to compute the current in response to different driving frequencies.
To further confirm that the discretization did not introduce artifacts, we repeated those calculations with Langevin simulations.
Agreement is clear from Fig.~\ref{langvtilt}, which shows negative currents at low frequency, small positive currents at high frequency, and a subtle current reversal around \(f \equiv 1 / \tau = 1100 \si{\kHz}\).
We focused on the jump process model to understand the nature of the current reversal: why does current vanish in the limit of infinitely slow or infinitely fast driving, why does low-frequency driving push particles to the left while high-frequency pushes to the right, what sets the frequency scale of the crossover, and how could the potential be sculpted so as to make the current reversal more pronounced?

The first question is the most straightforward.
At low frequencies, the driving is slow enough that the system can equilibrate on each landscape before the drive toggles to the other landscape.
Because equilibrium systems do not support currents, the transient flow developed upon switching between rate matrices \(\mathsf{W}_1\) and \(\mathsf{W}_2\) is the only source of current. 
The period-averaged current, which bears a prefactor \(\tau^{-1}\), thus vanishes in the limit of large \(\tau\).

At high frequencies, the driving is so rapid that the system cannot respond fast enough to each segment of the driving potential; instead, the system feels an averaged rate matrix \(\mathsf W_\text{eff}=(\mathsf{W}_1 + \mathsf{W}_2) / 2\).
In other words, there is a separation of timescales between that of the driving potential and that of the system's response. 
Mathematically, this intuition follows from the high-frequency \((\tau \to 0)\) limit of the Baker-Campbell-Hausdorff expansion
\begin{equation}
\exp\left[\frac{\tau \mathsf{W}_2}{2}\right] \exp\left[\frac{\tau \mathsf{W}_1}{2}\right] = \exp\left[\tau\mathsf{W}_\text{eff} + O(\tau^2)\right].
\end{equation}
Because both \(\mathsf{W}_1\) and \(\mathsf{W}_2\) are derived from potential energy landscapes, they both obey detailed balance.
Their average, which is associated with the average of the two potential energy landscapes, must likewise obey detailed balance and have vanishing current.

Between the \(f \to 0\) and \(f \to \infty\) extremes, the current depends on the kinetics of driven barrier crossing events, which cannot be so easily rationalized.
Spatial discretization and associated spectral methods offer a powerful tool to numerically interrogate the intermediate regime without the noise of trajectory sampling.

\subsection{Origin of current reversal}
\label{sec:origin}
\begin{figure*}[htb]
    \centering
    \begin{tikzpicture}
        \node at (0,6.4) {\includegraphics[width=0.45\textwidth]{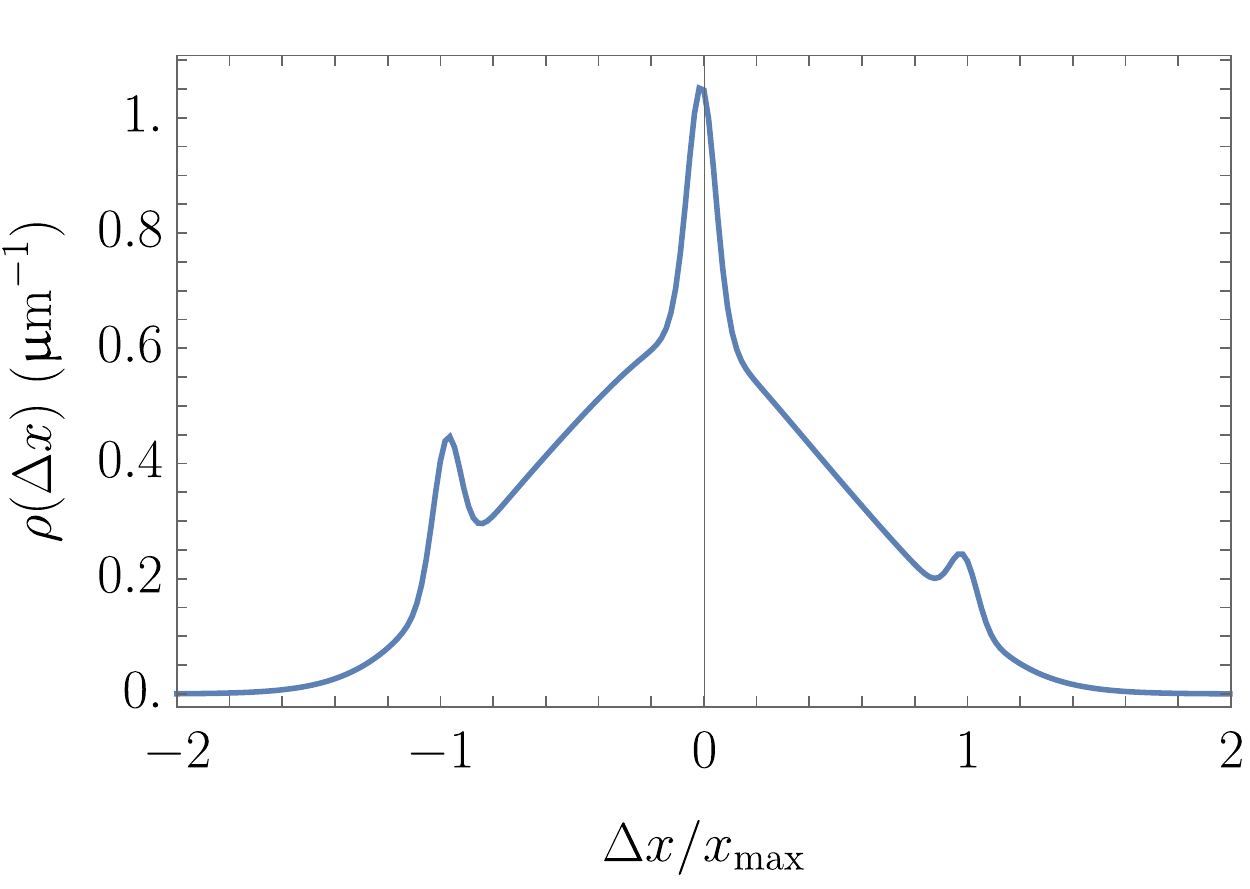}};
        \node at (8.5,6.4) {\includegraphics[width=0.49\textwidth]{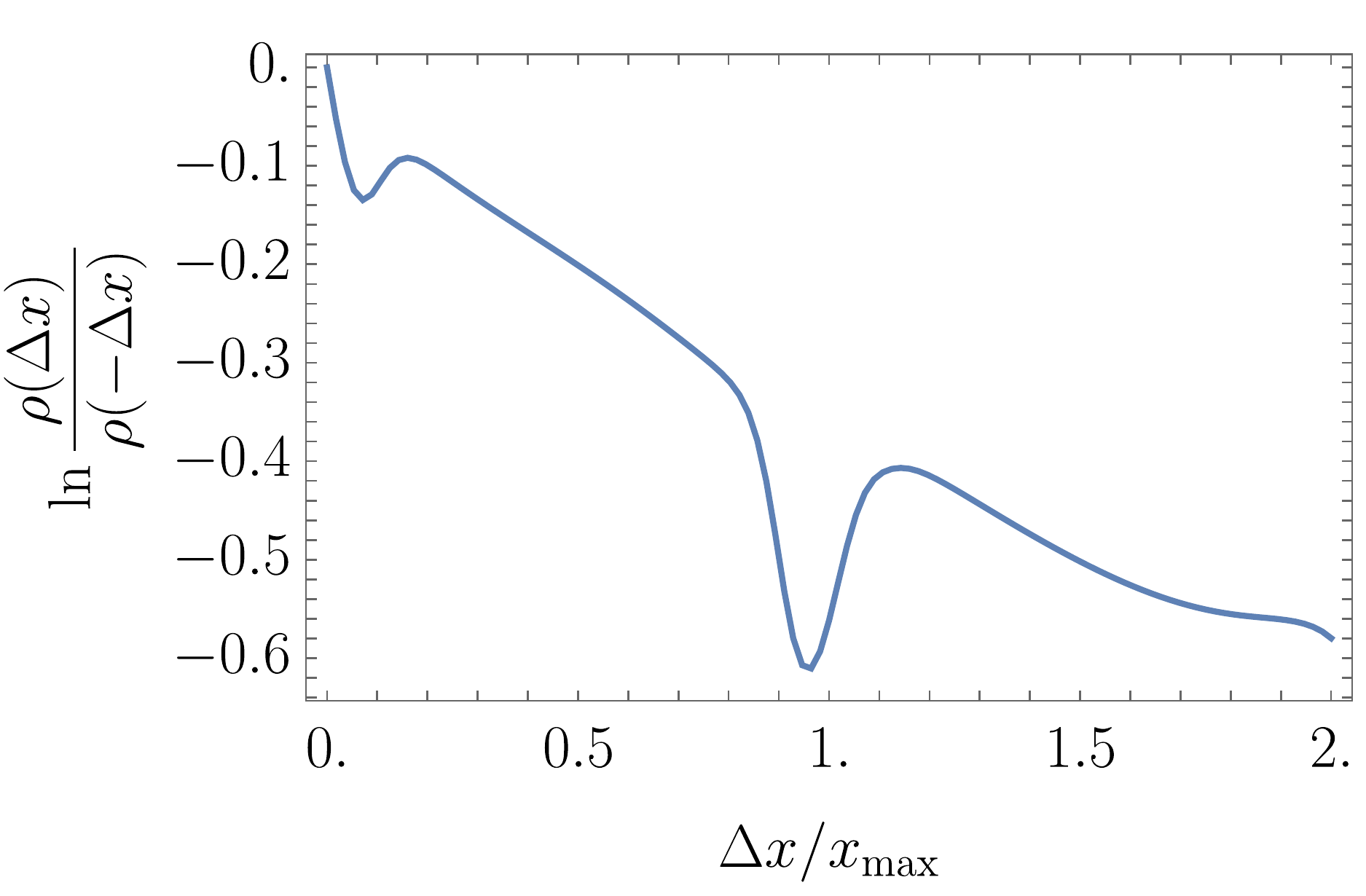}};
        \node at (0,0.7) {\includegraphics[width=0.45\textwidth]{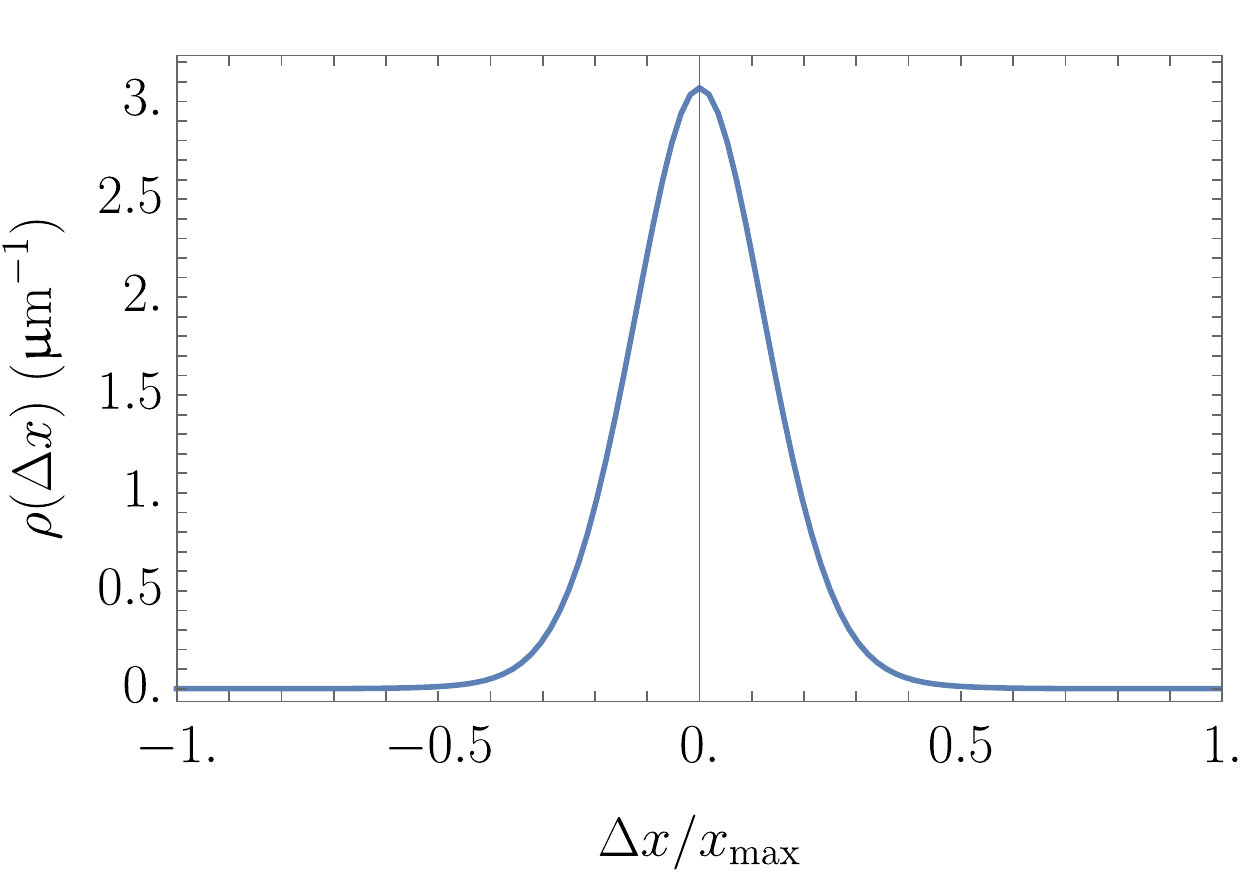}};
        \node at (8.5,0.7) {\includegraphics[width=0.49\textwidth]{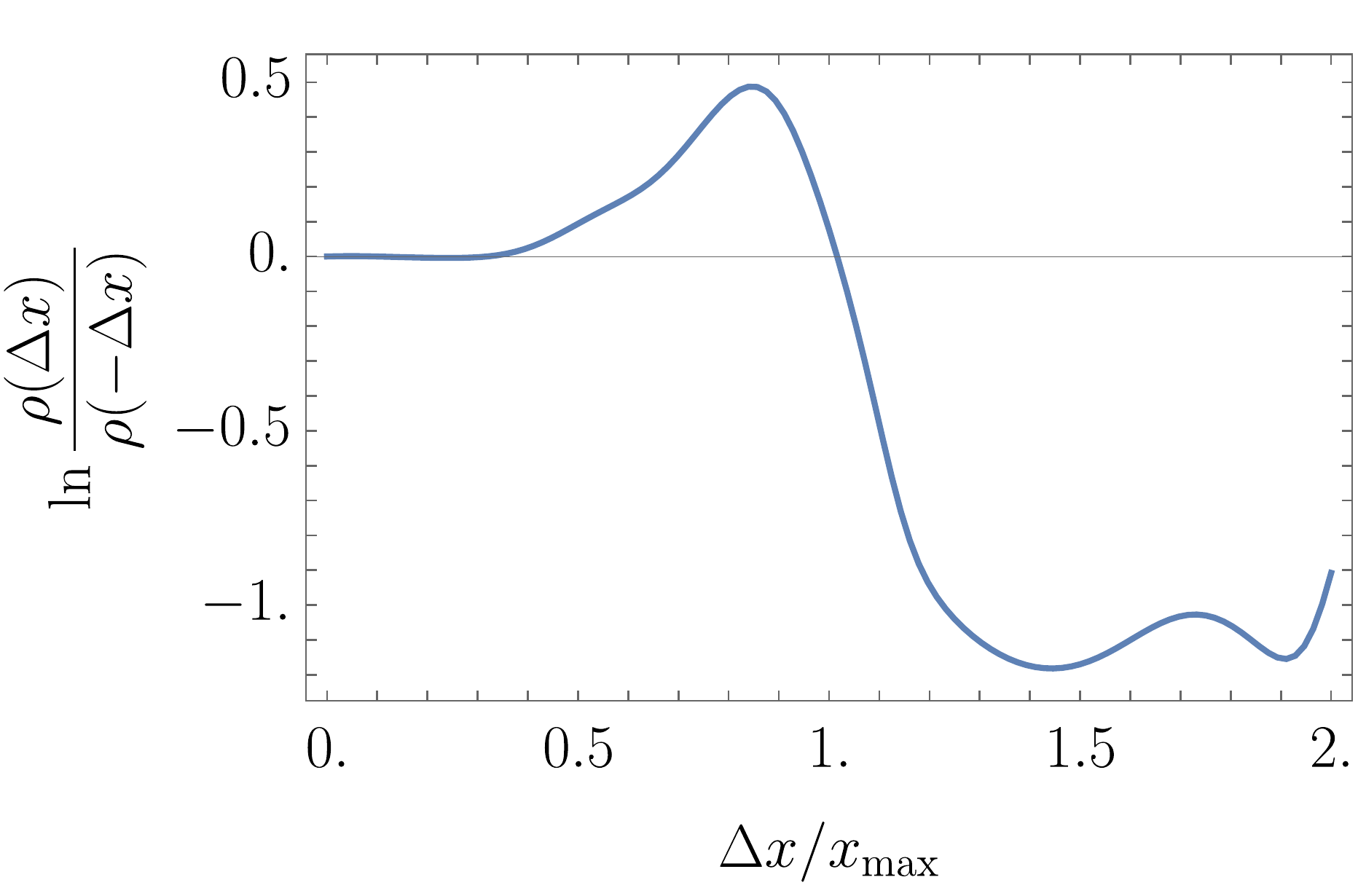}};
        \node at (-2.5,8.6) {(a)};
        \node at (6.6,8.6) {(b)};
        \node at (-2.5,2.9) {(c)};
        \node at (6.6,2.9) {(d)};
    \end{tikzpicture}
    \caption{Asymmetry of single-period displacements.
      After a full driving period of length \(\tau\), the jump process results in a displacement \(\Delta x\).
      (a) At low frequency, i.e., 100~\si{\kHz}, this displacement distribution is dominated by three peaks which stay in the original potential minima or move to adjacent minima to the left or right.
      (b) The imbalance between right and left, reported by \(\ln[\rho(\Delta x)/\ln \rho(-\Delta x)]\), shows a low-frequency preference for leftward currents regardless of the magnitude of displacement.
      By contrast, (c) while displacements at high frequency, i.e. 1600~\si{\kHz}, are nearly symmetrical and noticeably smaller than the spatial period \(x_{\rm max}\), (d) a left-right asymmetry can be seen for the rare large displacements whose magnitude exceeds half the spatial period.
      All distributions are computed using a grid spacing \(h = 1 / 56~\si{\um}\).
}
    \label{trajprob}
\end{figure*}
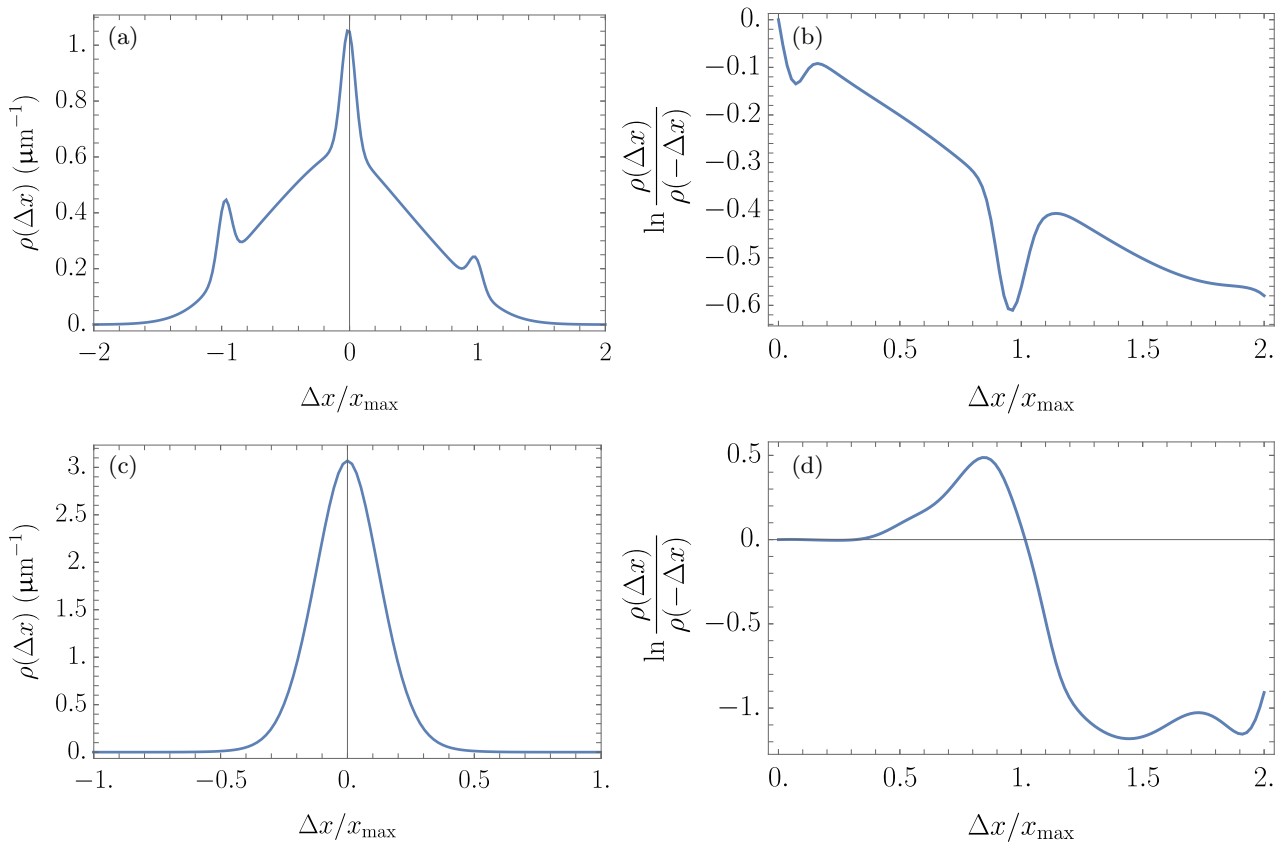

\begin{figure*}[htb]
    \centering
    \begin{tikzpicture}
        \node at (0,6.4) {\includegraphics[width=0.38\textwidth]{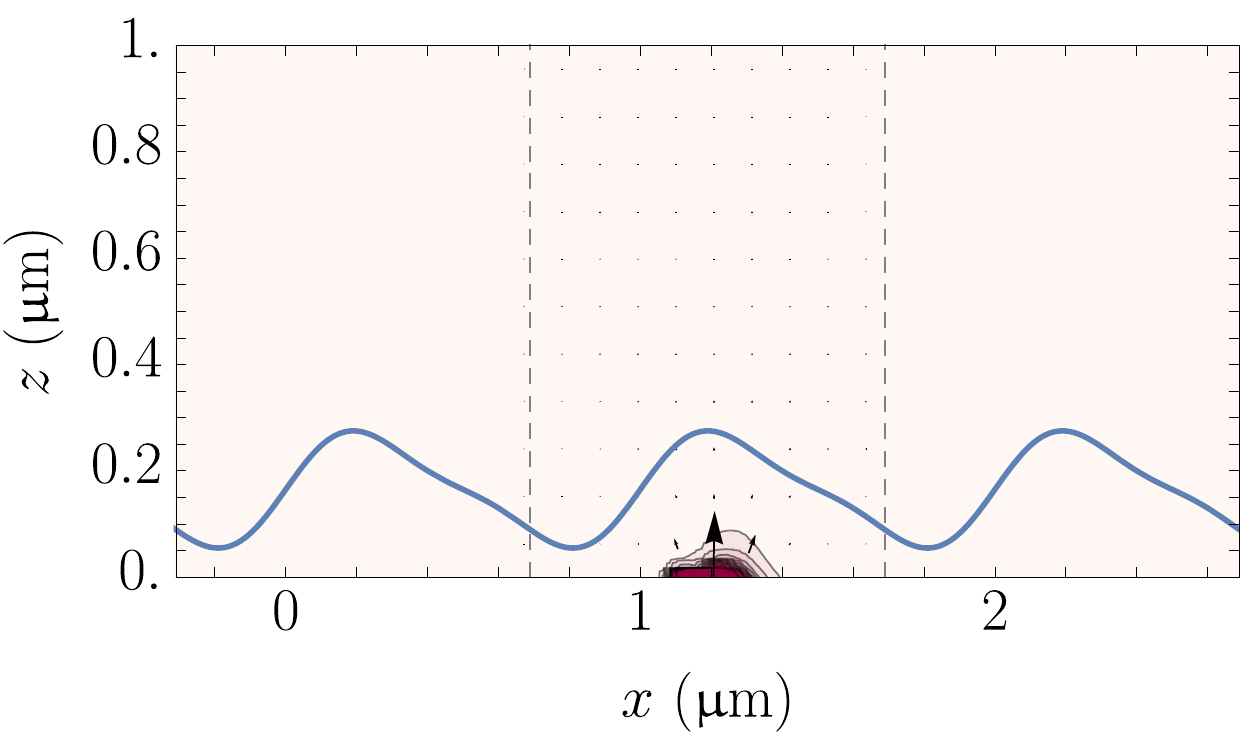}};
        \node at (7.6,6.4) {\includegraphics[width=0.38\textwidth]{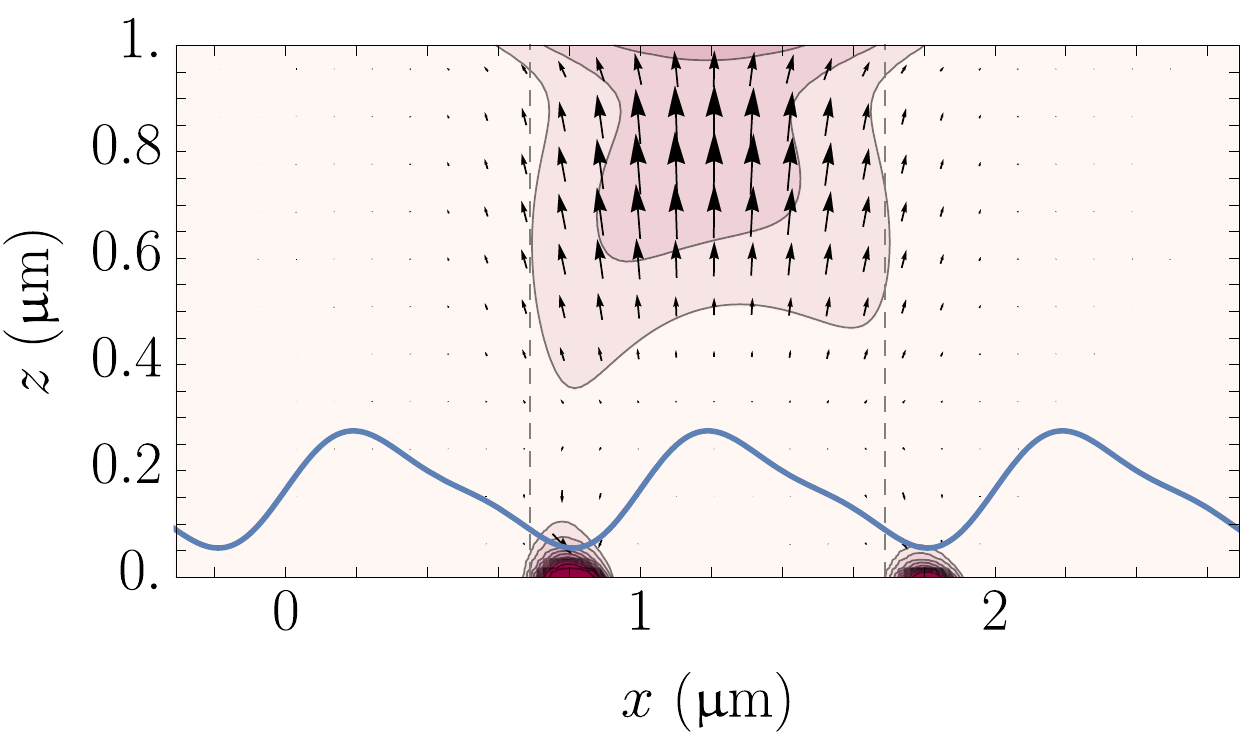}};
        \node at (0,2) {\includegraphics[width=0.38\textwidth]{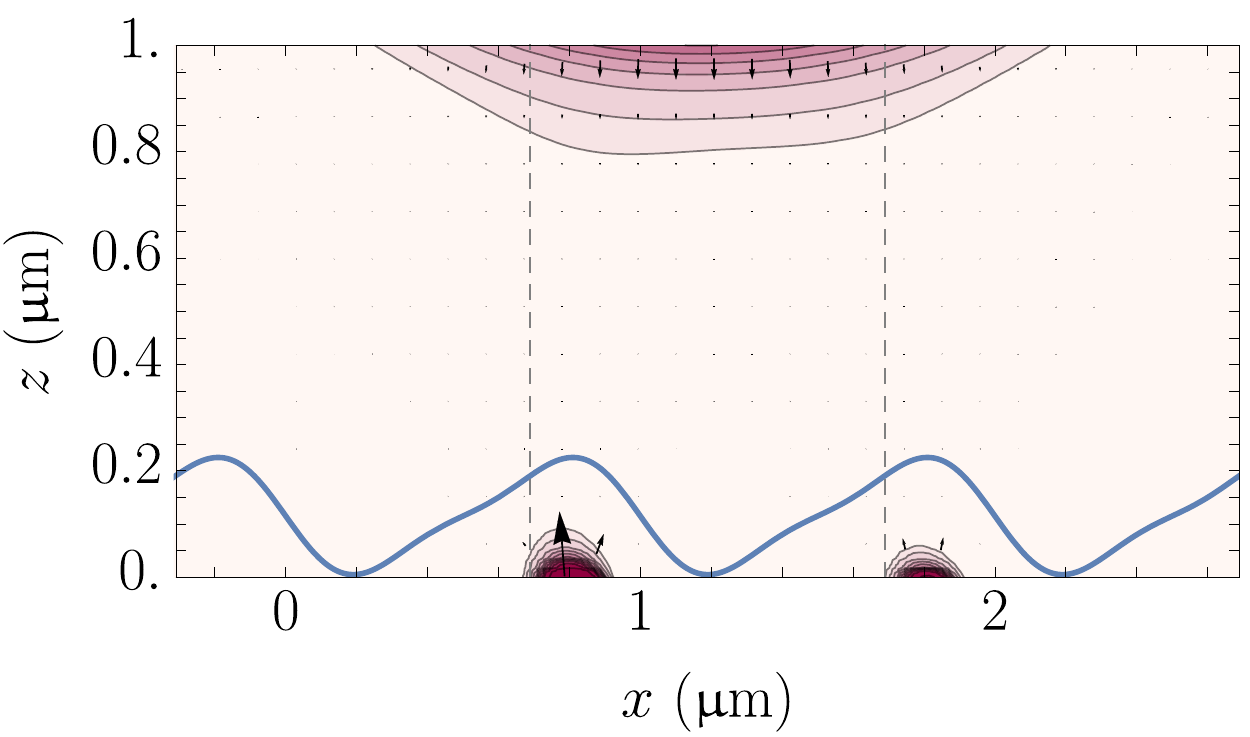}};
        \node at (7.6,2) {\includegraphics[width=0.38\textwidth]{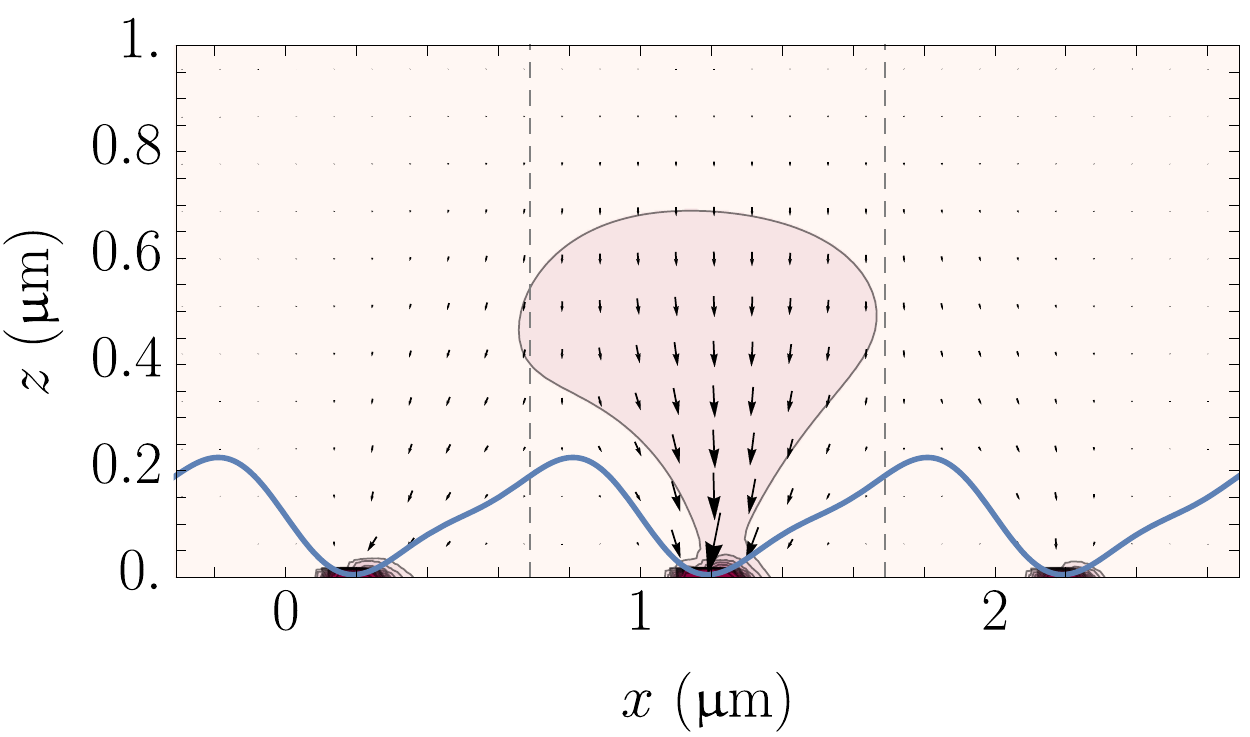}};
        \node at (-5,4.4) {\includegraphics[width=0.17\textwidth]{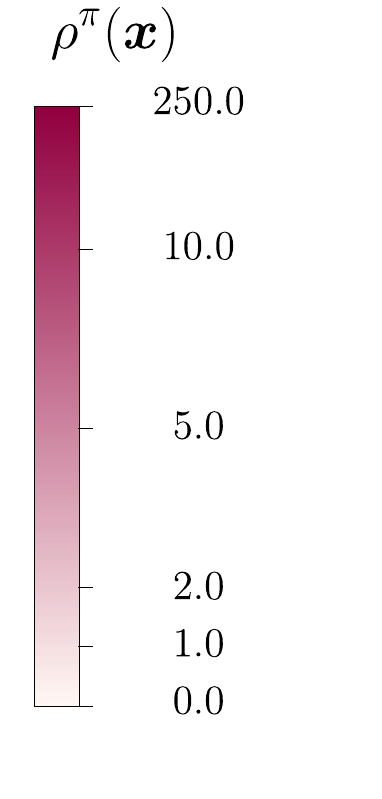}};
        \node at (-2.1,7.9) {(a)};
        \node at (5.5,7.9) {(b)};
        \node at (-2.1,3.5) {(c)};
        \node at (5.5,3.5) {(d)};

    \end{tikzpicture}
    \caption{Evolution of density at low frequency.
Snapshots of densities and currents obtained by spectral calculations for a single particle hopping across multiple periodic replicas with driving frequency \(f_{\rm lf}=100~\si{\kHz}\).
Densities are plotted with a hyperbolic tangent scale to emphasize the regions of low probability.
The central replica (enclosed by dashed lines) is initialized at \(t=0\) in the time-periodic steady-state distribution.
Each blue curve across the bottom of a plot is the \(z=0\) cross section of the potential \(U\) upon which the dynamics will relax for the time \(\tau / 4\) that separates each snapshot: (a) \(t=0\), (b) \(t=\tau/4\), (c) \(t=\tau/2\), and (d) \(t=3\tau/4\).
That potential, with peak-to-trough height of \(V_{\rm max}\), is scaled and shifted to rest along \(z = 0\) in the plot to highlight the \(x-\)dependence of the potential along the bottom surface.
All distributions are computed using a grid spacing \(h = 1 / 56~\si{\um}\).
}
    \label{extdens}
\end{figure*}
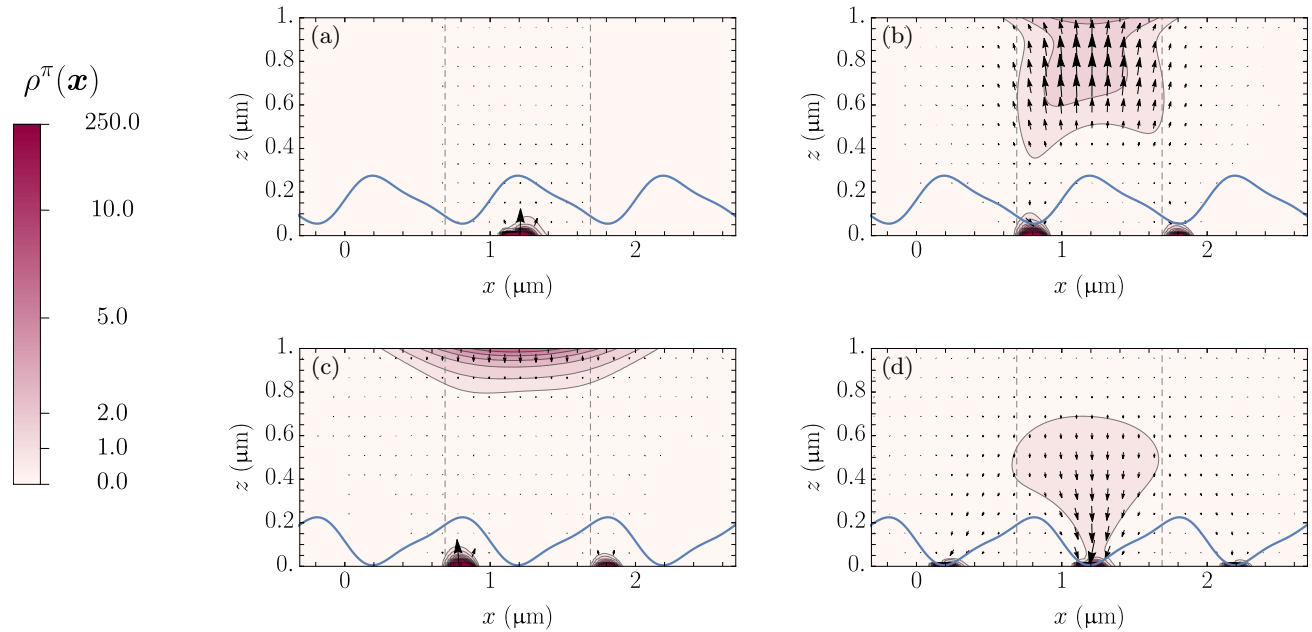

To compare the low- and high-frequency behaviors, we focused on the driving frequencies that give rise to the maximum and minimum currents from Fig.~\ref{langvtilt}.
We will consider a characteristic low-frequency driving to be \(f_{\rm lf} = 100~\si{\kHz}\) and a characteristic high-frequency driving to be \(f_{\rm hf} = 1600~\si{\kHz}\), chosen to roughly correspond to frequencies resulting in the most negative and most positive currents, respectively.
As described in Sec.~\ref{sec:spectral}, we computed the time-periodic steady-state distributions on the grid for each driving frequency.
Those steady-state distributions were used as the initial density, which was then propagated for a full period.
To keep track of motion into the neighboring replicas, the periodic boundary conditions were unfolded and calculations were performed using a collection of five neighboring cells surrounded by closed boundaries~\footnote{Five cells were sufficient to replace the periodic boundary conditions because, with overwhelming probability, a particle moves no further than the neighboring cell in a single period of driving.}.
After initializing density in the central cell (see Fig.~\ref{extdens} for an illustration of the central three replicas), the net displacement along the \(x\) direction was computed with an explicit matrix propagator.
The distribution of this displacement, \(\rho(\Delta x)\), is plotted in Fig.~\ref{trajprob}, showing that low-frequency displacements are dominated by shifts of the form \(\Delta x = n x_{\rm max}\) for \(n = 0, \pm 1\) with a net leftward preference.
In contrast, high-frequency displacements more closely resemble thermal motion---the displacements have a nearly Gaussian distribution about \(\Delta x = 0\).
The high-frequency currents seem to emerge from a subtle asymmetric deviation from normality.

To more clearly illuminate the asymmetry in both displacement distributions, we also plot in Fig.~\ref{trajprob} the relative probability of \(+\Delta x\) and \(-\Delta x\) as a function of the magnitude of displacement.
This plot shows a low-frequency asymmetry for all values of \(\Delta x\), but the high-frequency displacement distribution appears to be symmetric up to a length scale of about \(x \sim x_{\rm max} / 2\).
Notably, displacements of this magnitude or larger are exceedingly rare when \(f = f_{\rm hf}\).
In other words, the low-frequency asymmetry is present for typical trajectories, whereas the high-frequency asymmetry emerges only at the level of rare events.

To gain a mechanistic perspective into that difference, we traced the time-dependent flow of probability, starting with the low-frequency case.
In Eq.~\eqref{eq:interiorprobs}, we had computed the time-periodic steady-state density from a top eigenvector.
We propagated this density over one temporal period using the rate matrix \(\mathsf{W}\) that imposed periodic boundary conditions.
To most simply distinguish between leftward and rightward currents, we then propagated this density without periodic boundaries.
That evolution of density, shown in Fig.~\ref{extdens}, reflects a mechanism reminiscent of a one-dimensional flashing ratchet.
The motion in the \(x\) direction switches between a sawtooth and a flat potential, with the switch triggered by periodic motion along the \(z\) direction.

The periodic motion in the \(z\) direction is easily rationalized from the tilting of the landscapes in Fig.~\ref{twopots}, and the resulting current along \(x\) follows from the flashing ratchet mechanism.
Particles that accumulate near \(z_{\rm max}\) at \(t = \tau / 2\) can diffuse left or right symmetrically, but as the particles descend toward \(z = 0\) during the \(\tau / 2 \leq t \leq \tau\) relaxation, more of the leftward diffusing particles will have made it past the barrier than their rightward counterparts.
The result is net motion that yields a displacement by one spatial period to the left more often than to the right.
Thus, the direction of low-frequency motion requires one to inspect the shape of the \(z = 0\) potential (blue lines in Fig.~\ref{extdens}) during the relaxation from high to low \(z\) \((\tau / 2 \leq t \leq \tau)\); current will move along \(x\) in the direction with the shortest trough-to-peak distance as this will be the barrier around which it is easiest to diffuse.

\begin{figure*}[htb]
    \centering
    \begin{tikzpicture}
        \node at (-1.9,-2.4) {\includegraphics[width=0.17\textwidth]{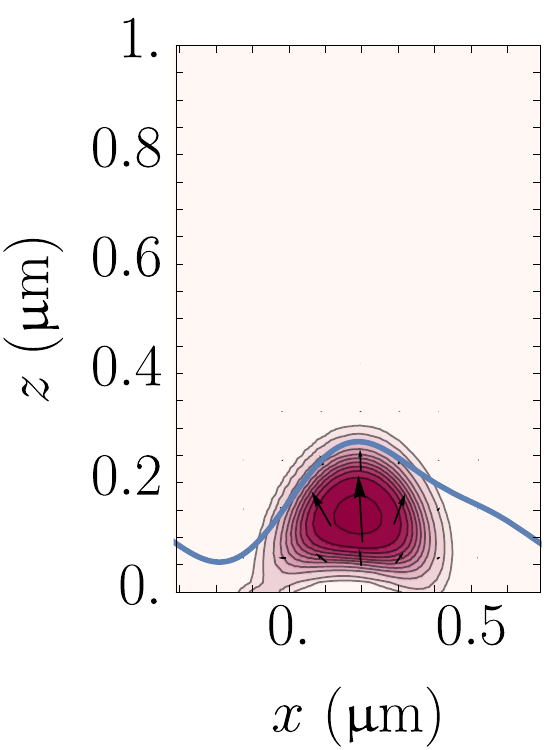}};
        \node at (1.9,-2.4) {\includegraphics[width=0.17\textwidth]{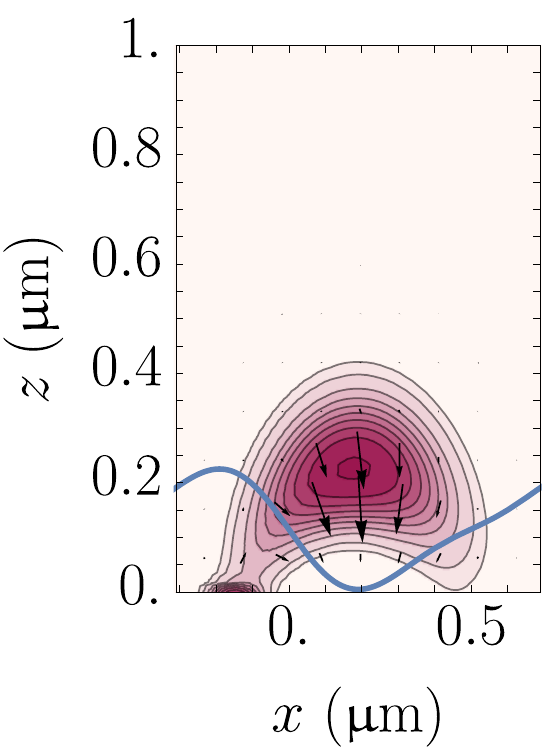}};
        \node at (5.7,-2.4) {\includegraphics[width=0.17\textwidth]{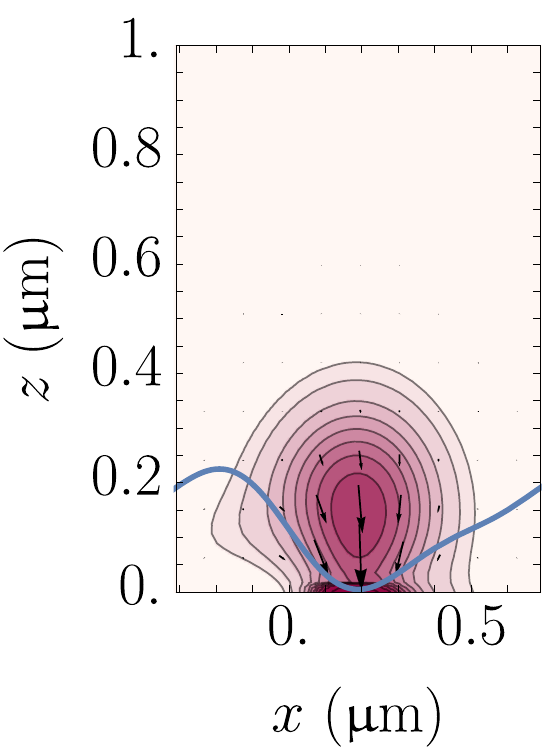}};
        \node at (9.5,-2.4) {\includegraphics[width=0.17\textwidth]{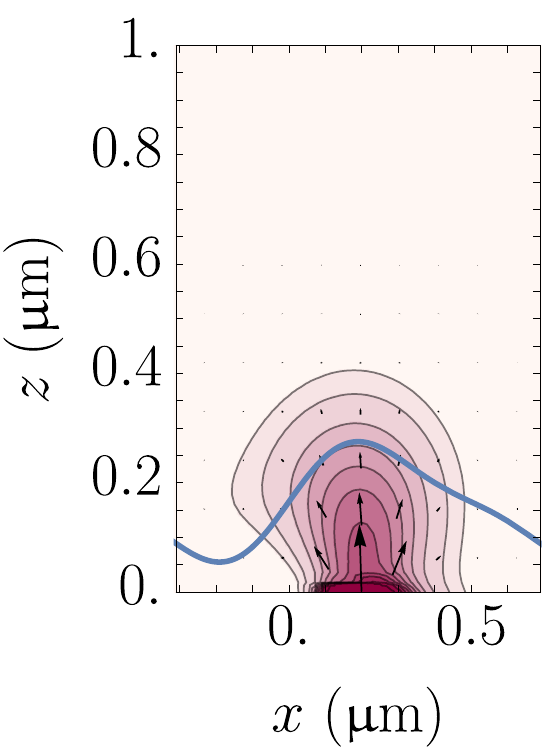}};
        \node at (-1.9,-6.8) {\includegraphics[width=0.17\textwidth]{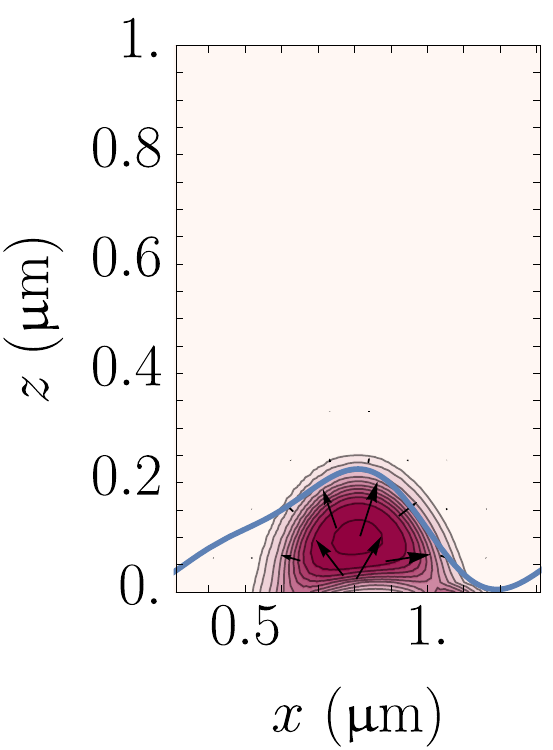}};
        \node at (1.9,-6.8) {\includegraphics[width=0.17\textwidth]{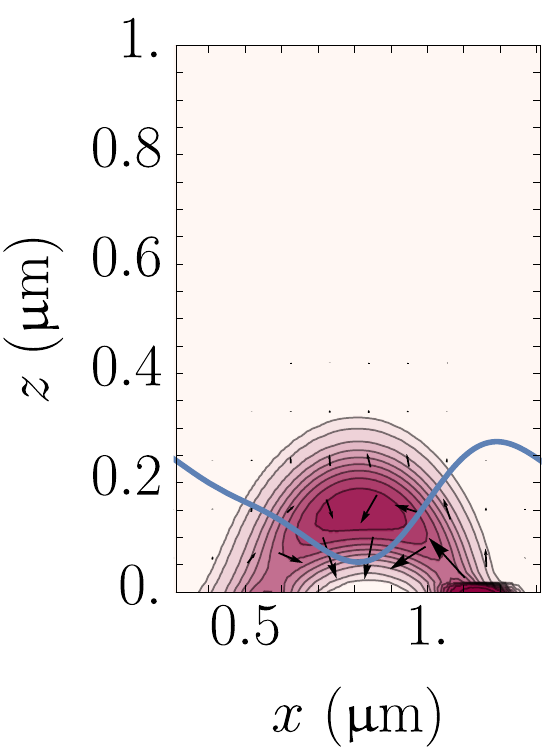}};
        \node at (5.7,-6.8) {\includegraphics[width=0.17\textwidth]{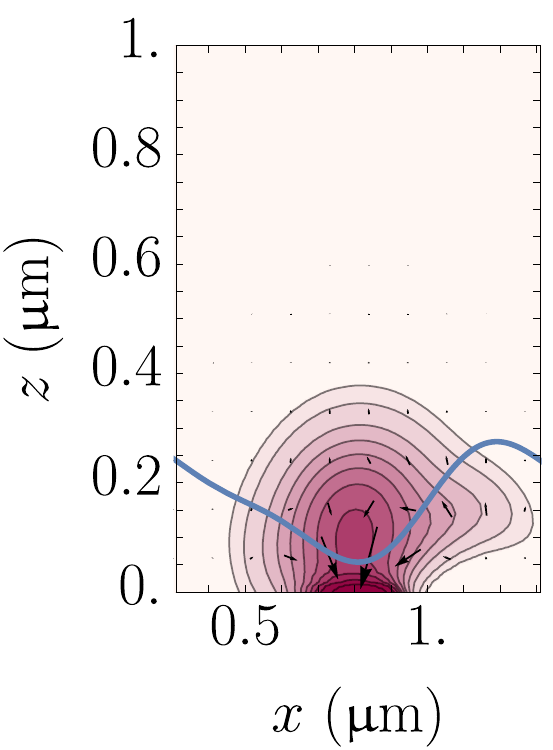}};
        \node at (9.5,-6.8) {\includegraphics[width=0.17\textwidth]{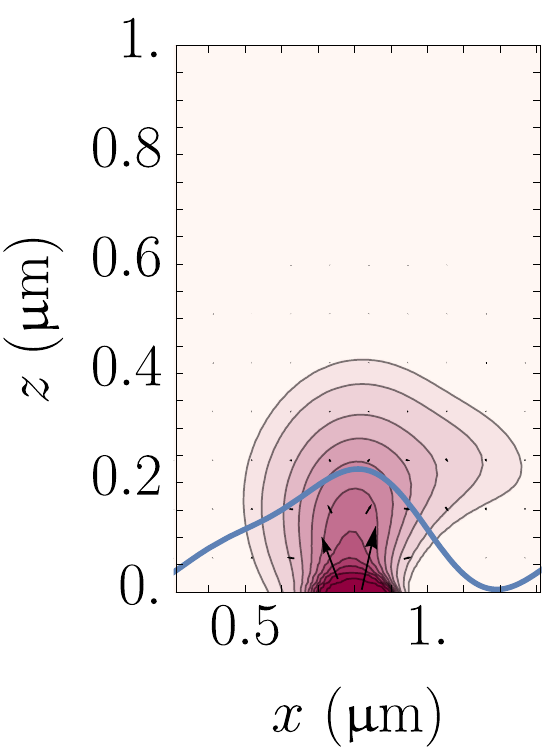}};
        \node at (-5,-3.8) {\includegraphics[width=0.17\textwidth]{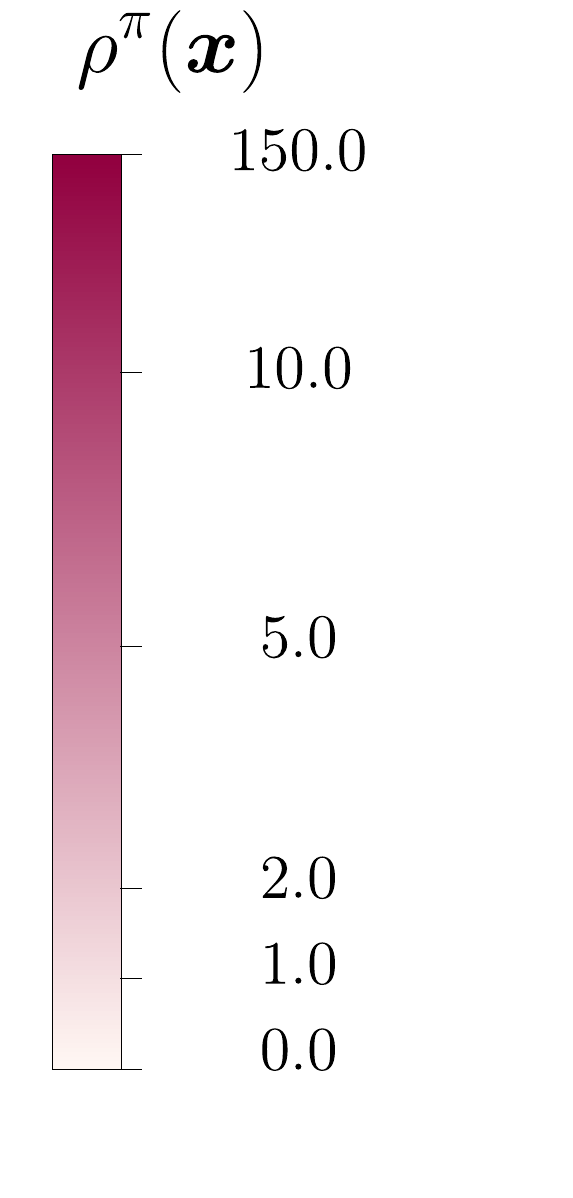}};

        \node at (-2.1,-0.85) {(a)};
        \node at (1.7,-0.85) {(b)};
        \node at (5.5,-0.85) {(c)};
        \node at (9.3,-0.85) {(d)};
        \node at (-2.1,-5.25) {(e)};
        \node at (1.7,-5.25) {(f)};
        \node at (5.5,-5.25) {(g)};
        \node at (9.3,-5.25) {(h)};
    \end{tikzpicture}
    \caption{Asymmetry in the evolution of density at high frequency.
      Densities are plotted with a hyperbolic tangent scale to emphasize the regions of low probability.
      The relaxation of density and current is depicted at (a) \(t=\tau/4\), (b) \(t=\tau/2\), (c) \(t=3\tau/4\), and (d) \(t=\tau\), when initialized at the bottom of the well at time 0.
      Snapshots of the density and current are also shown at (e) \(t=\tau/4\), (f) \(t=\tau/2\), (g) \(t=3\tau/4\), and (h) \(t=\tau\), when initialized at the trough at time \(\tau / 2\).
      Because the density is slightly repelled from \(z = 0\) in (a) and (b) compared to (e) and (f), much less density moves far enough left to reach the \(z = 0\) well in (b) than the density that moves right in (f).
      This imbalance results in the rightward fat tail in (h) that exceeds the leftward fat tail in (d).
      As in Fig.~\ref{extdens}, a lattice with grid spacing \(h = 1 / 56~\si{\um}\) was employed, and the blue curve across the bottom of the plots is the \(z=0\) cross section of the potential \(U\) upon which the dynamics will relax for the time \(\tau / 4\) that separates each snapshot.}
    \label{extdens2}
\end{figure*}
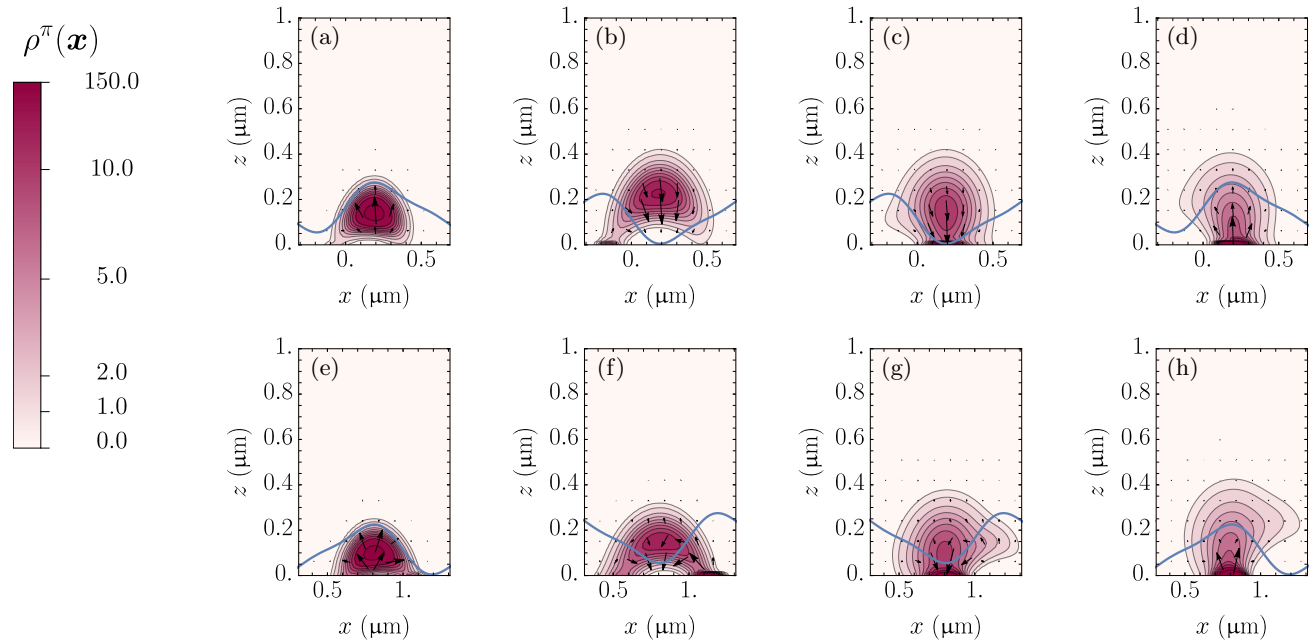

It is more difficult to identify the mechanism for high-frequency positive current because the displacement distributions are nearly symmetrical.
The asymmetry is more subtle than in the low-frequency case, so it cannot be simply observed in plots similar to Fig.~\ref{extdens}.
Instead, we detect the origin of the asymmetry by studying the evolution from an initial condition atop a local maximum of the energy.
Relaxation from this initial condition is particularly revealing since the probability that accumulates in a trough at the end of one period is situated near the local maximum as the next period begins.
One can therefore reason that the dominant trajectories are those which relax from a local maximum, but there are two such maxima: one at \(t = 0\) and another at \(t = \tau / 2\), shown in the two rows of Fig.~\ref{extdens2}.
The period-averaged current arises out of the balance of those two relaxation processes.

If not for motion along \(z\), symmetry arguments would require the two relaxations to be mirror images of each other, yielding symmetric displacement distributions and vanishing current.
But the two rows of Fig.~\ref{extdens2} are not mirror images, and the imbalance between the probability of the large displacements in the ``fat tails'' of the distributions explains the net positive current.
That imbalance of the large displacements is traced back to a slight difference in the accumulation of probability in the \(z = 0\) wells shown in Figs.~\ref{extdens2}(b) and \ref{extdens2}(f).
Due to the tilt of the potential along \(z\), Fig.~\ref{extdens2}(b) has less probability accumulate in its well than Fig.~\ref{extdens2}(f).
Any displacements that are too small to transit from peak to well (\(x < x_{\rm max} / 2\)) are not appreciably affected by the difference, but the rare trajectories that move all the way from peak to well are thus more favored in Fig.~\ref{extdens2}(f) when the tilt steers particles toward \(z = 0\).
Therefore the effect of the tilt is to favor the peak-to-trough motion during the \(\tau / 2 \leq t \leq \tau\) relaxation over the complementary motion during \(0 \leq t \leq \tau / 2\), yielding net positive current.

The importance of the rare large displacements explains why the direction of high-frequency motion is exactly opposite that of low-frequency motion.
For the low-frequency case, we already noted that the direction of motion could be simply explained by the shortest \emph{trough-to-peak} distance of the applied potential at \(z = 0\) during the \(\tau / 2 \leq t \leq \tau\) relaxation.
At high frequency, current instead moves along \(x\) in the direction with the shortest \emph{peak-to-trough} distance.

\subsection{Sculpting the energy landscapes}

We have traced a reversal in the sign of the steady-state horizontal current back to the competition between two opposing classes of mechanisms.
However, the current reversal is subtle for the potential \(U\)---any current generated under driving frequencies past 1100~\si{\kHz} is almost imperceptible.
How can one sculpt the energy landscape such that both negative and positive currents are similar in magnitude?
We show that the low- and high-frequency mechanisms discussed in Sec.~\ref{sec:origin} inform the manner in which the landscape should be altered.

\begin{figure*}[htb]
    \centering
    \begin{tikzpicture}
        \node at (0,6.4) {\includegraphics[width=0.45\textwidth]{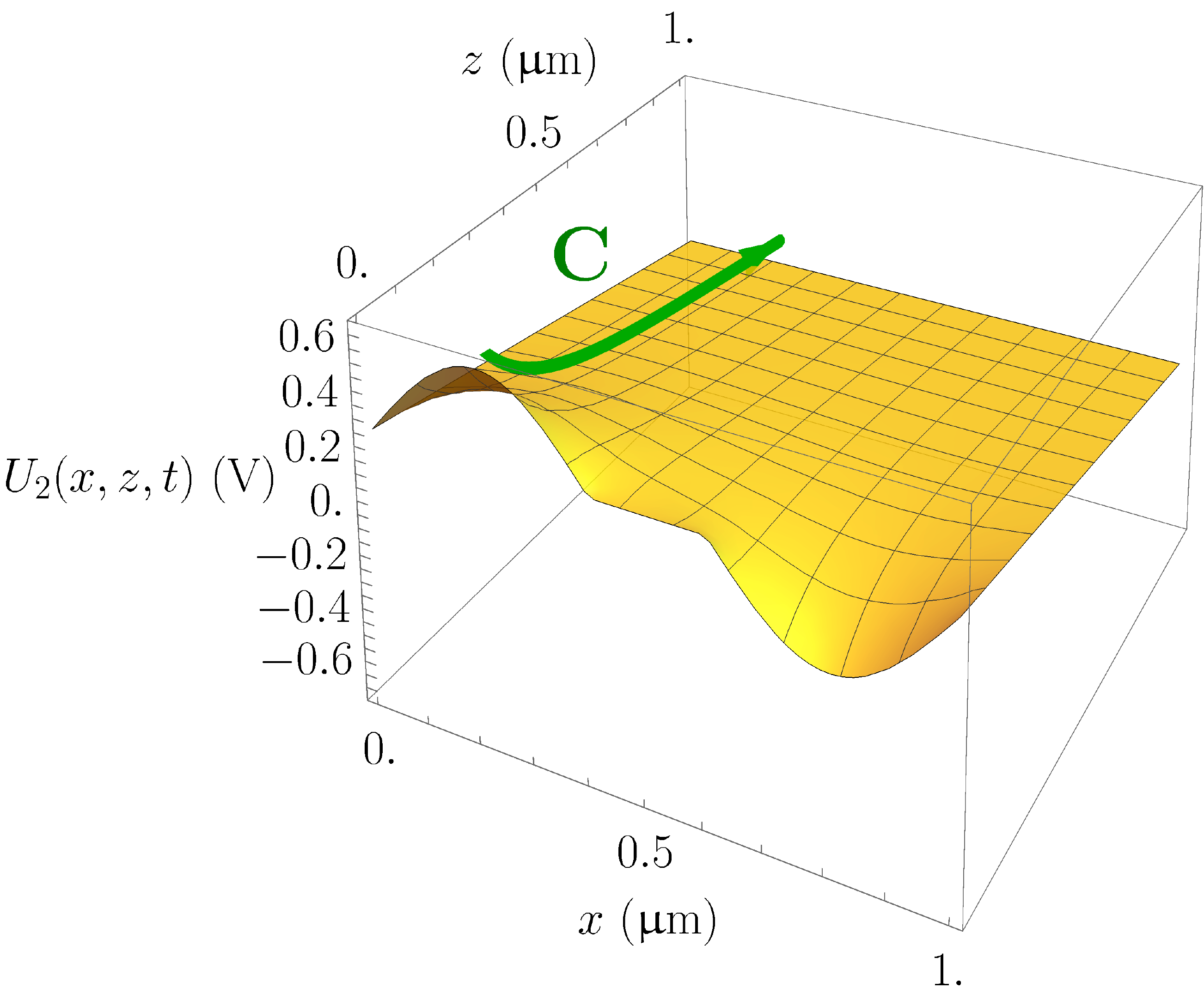}};
        \node at (8.5,6.4) {\includegraphics[width=0.45\textwidth]{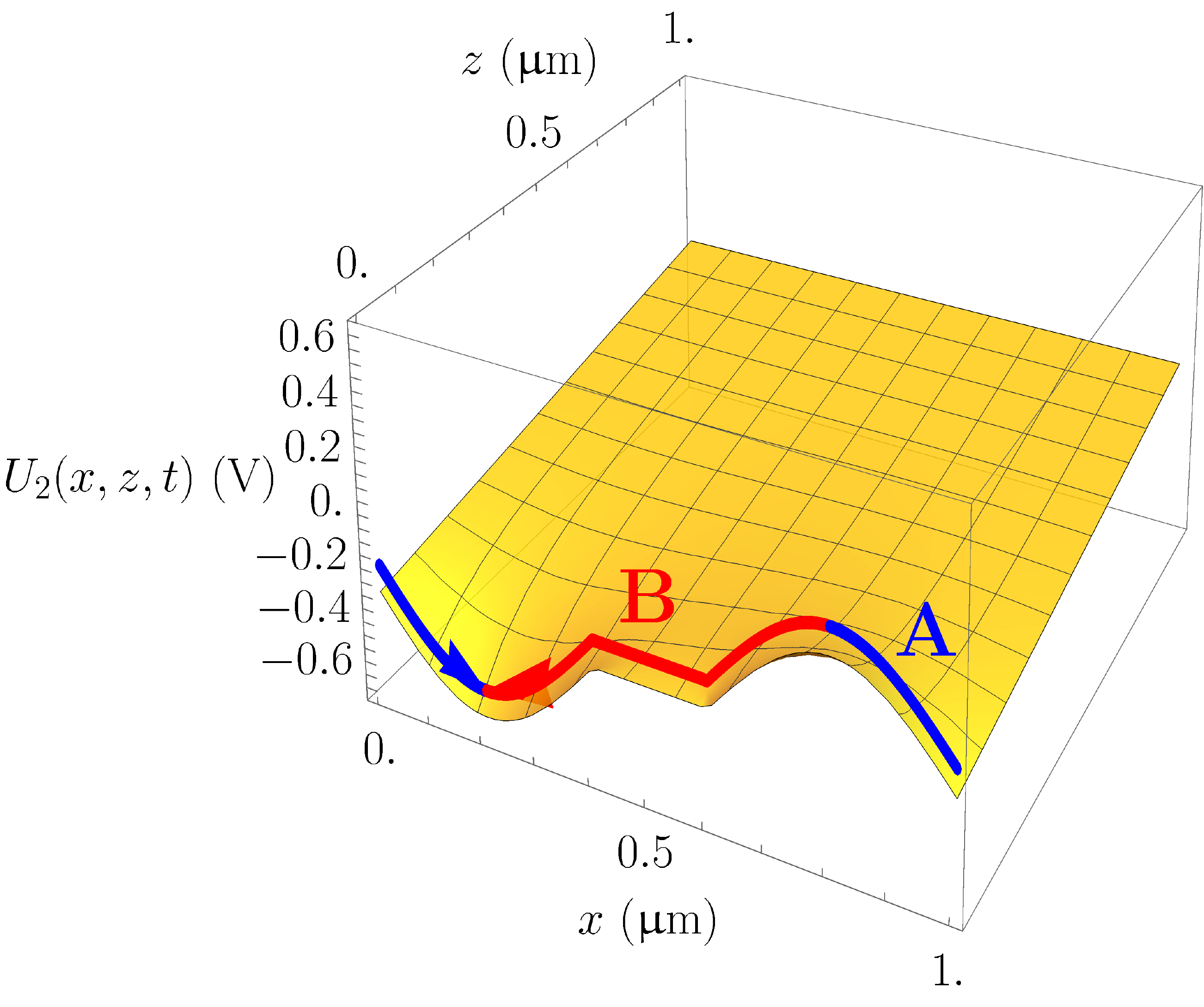}};
        \node at (-2,8.8) {(a)};
        \node at (6.5,8.8) {(b)};
    \end{tikzpicture}
    \caption{Two-dimensional landscapes associated with the electrostatic potential \(U_2\) at times (a) \(0<t\leq\tau/2\) and (b) \(t>\tau/2\). 
The arrows correspond to pathways that regulate the current reversal.
Timescales for one-dimensional diffusion along pathways A, B, and C are reported in Table~\ref{timescales}.}
    \label{otherpot}
\end{figure*}

To make the current switch between positive and negative values of similar magnitude, we seek a potential that curtails the low-frequency negative current and enhances the high-frequency positive current.
We developed intuition about how to achieve this goal by focusing on one-dimensional diffusion along three different pathways, labeled A, B, and C on the new sculpted landscape in Fig.~\ref{otherpot}.
The new electrostatic potential \(U_2(x,z,t)\) is computed from a numerical solution to Laplace's equation on a 150-by-150 mesh with boundary conditions as in \(U(x,z,t)\), but with spatial component
\begin{equation}
    X(x)=\begin{cases}
    \dfrac{1}{2}+\dfrac{a_3}{2}\sin\left(\dfrac{5\pi x}{2x_\text{max}}\right), & r\leq\frac{2}{5},\\
    \dfrac{1}{2}, & \frac{2}{5}<r\leq\frac{3}{5},\\ 
    \dfrac{1}{2}-\dfrac{a_3}{2}\cos\left(\dfrac{5\pi x}{2x_\text{max}}\right), & r>\frac{3}{5},
    \end{cases}
    \label{eq:Xx2}
\end{equation}
in terms of \(r=x/x_\text{max}-\lfloor x/x_\text{max}\rfloor\).
The parameter \(a_3\) is assigned the value 1.1 to match the amplitude of potential \(U\) at \(z=0\).
The new potential was designed to have a plateau in \(X(x)\) so that the diffusion along A remains essentially unaffected, but the timescale for motion along B will increase.
Moreover, the constant vertical offset at \(z=0\) being smaller in \(U_2\) than in \(U\) increases the timescale for motion along C as was confirmed by Gillespie simulations of one-dimensional diffusion on both the old (\(U\)) and new (\(U_2\)) landscapes.
The first-passage times along those pathways, collected in Table~\ref{timescales}, confirm that the changes in the landscape have the desired effect of slowing diffusion along B and C.
Though we did not intend to appreciably alter the timescale for diffusing along A, that diffusion was slightly faster on landscape \(U_2\) than on the original landscape \(U\).

\begin{table}[htb]
  \begin{tabular}{lll}
    \hline
    Pathway & \(U\)-driven (\si{\ns}) & \(U_2\)-driven (\si{\ns}) \\
    \hline\hline
    \textcolor{blue}A & \(561\pm2\) & \(518\pm2\) \\
    \textcolor{red}B & \(1087\pm3\) & \(2408\pm15\) \\
    \textcolor{green!50!black}C & \(3613\pm13\) & \(4324\pm17\) \\
    \hline
  \end{tabular}
  \caption{Timescales for diffusion along pathways A, B, and C of Fig.~\ref{otherpot} for potential landscapes \(U\) and \(U_2\).
Reported timescales are the averages and standard errors of the first-passage time for traversing the pathways, collected from \(10^4\) one-dimensional Gillespie simulations during relaxation on a fixed landscape.
Relaxation from peak to trough along the two horizontal pathways, A and B, was simulated using the \(\tau / 2 \leq t \leq \tau\) potential,d while C relaxed using the \(0 \leq t \leq \tau / 2\) potential.
For comparison, the maximal negative current is generated with a driving frequency that switches landscapes after about 5200 and 5900 \si{\ns} for \(U\) and \(U_2\), respectively, so there is sufficient time for diffusion along all three pathways.
By contrast, the maximal positive current is generated by a driving frequency that allows relaxation on \(U\) and \(U_2\) for only around 310 and 630 \si{\ns}, respectively.
Note that with landscape \(U_2\), that driving rate provides enough time for descents down A which are significantly less rare than they are on landscape \(U\).}
  \label{timescales}
\end{table}

\begin{figure}[htb]
    \centering
    \includegraphics[width=0.45\textwidth]{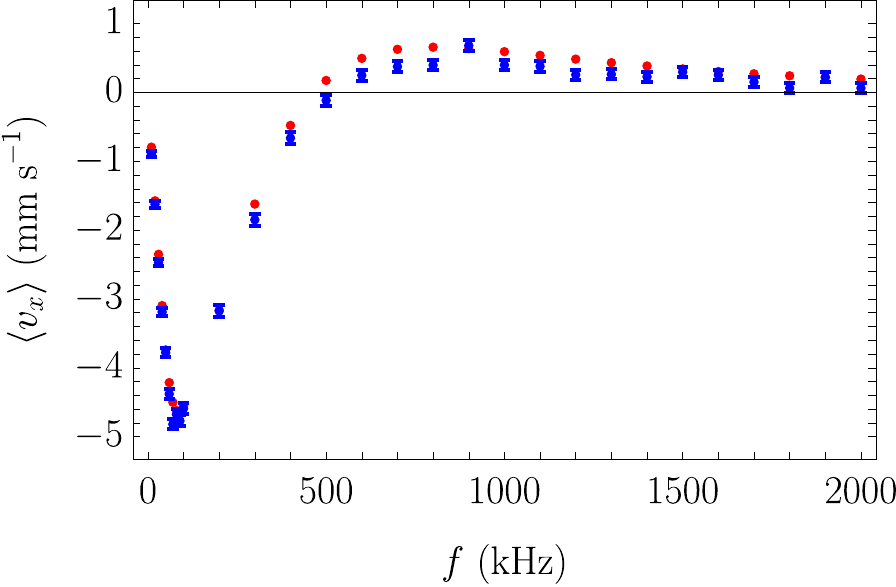}
    \caption{Frequency-dependent current with the altered landscape \(U_2\).
      Average horizontal particle velocities were computed for driving frequencies ranging from 10 to 2000~\si{\kHz} using potential \(U_2\) as the driving protocol.
      Currents were calculated using both spectral methods (red) and from single-particle Langevin simulations (blue) as described in the text.
      Spectral calculations were performed on a 100-by-101 grid.
      Langevin simulations were averaged over 512 independent 10~\si{\ms} trajectories with a time step \(\Delta t\) of 35~\si{\ps}.
      Relative to Fig.~\ref{langvtilt}, the current under this driving protocol exhibits stronger positive currents and weaker negative currents, thus rendering the current reversal more pronounced.
      Here, the crossover occurs at a frequency of approximately \(500~\si{\kHz}\), a value much smaller than that seen under driving with \(U\).}
    \label{langvtilt2}
\end{figure}

The changes to the motion along A, B, and C work in concert to enhance the rightward current (see Fig.~\ref{langvtilt2}).
Because the high-frequency trajectories that contribute rightward motion are rare, even a small speedup along A can render those critical large-deviation trajectories significantly less rare, thereby yielding more positive current at high frequency.
The increased positive current also arises by disfavoring the low-frequency mechanism.
The plateau along B provides a trap that catches some fraction of the trajectories that would have otherwise moved left along the flashing ratchet mechanism, and that mechanism becomes unfavorable at a lower current-reversal frequency due to the slower motion along C.

\section{Conclusions}
\label{sec:conclusions}

Thermalized equilibrium systems are fully characterized by the Boltzmann distribution.
If one aims to alter the steady state, it is sufficient to consider changes to the energy landscape without explicitly worrying about the dynamics on that landscape.
If, however, one switches between multiple energy landscapes, the ensuing nonequilibrium dynamics can relax into more complicated time-dependent steady states with ratcheting current.
In that event, the relaxation dynamics on the landscapes cannot be ignored.
In fact, it is the interplay between the timescale of this dynamics and the timescale of the switching landscapes that regulates current generation.
This additional complexity means that efforts to design a time-dependent landscape that generates a desired current will require explicitly modeling the dynamical system.

In this work, we discussed straightforward ways to model that dynamics---with discretization in time or in space.
Spatial discretization allowed us to view the problem as a Markov jump process, which could be conditioned to generate positive or negative current.
Analyses of the relative probability of those currents revealed that a leftward bias seen at low frequency results from an asymmetry in typical trajectories, whereas a rightward bias seen at high frequency stems from an asymmetry on the level of rare trajectories.
Furthermore, we showed how that insight allowed for modifications to the landscape that would impact the frequency dependence of the current.

Looking forward, it will be interesting to extend the tools and analysis in two complementary directions.
First, how well can one reverse-engineer energy landscapes given a desired frequency response as an input?
We have showed how to enhance the high-frequency positive current, but might it be possible to design more complex landscapes that support multiple current reversals?
The discrete-space lattice models of this work will provide a numerically efficient playground to explore how flexible of a frequency response is possible.
Second, while we have analyzed a single ratcheting particle, most real ratchets involve multiple interacting particles.
One route to considering the effects of interactions between particles is to add more particles to the time-dependent lattice models studied here.
In their simplest form, these models could be exclusion processes with time-dependent driving.
The spectral methods we employed would be challenged by the fact that the state space would grow exponentially with the number of particles, but the models could be analyzed using Gillespie simulations or potentially with approximations built on a matrix product ansatz.

\section{Acknowledgments}
The authors thank Hadrien Vroylandt, Ofer Kedem, and Emily Weiss for helpful discussions.
This research was supported in part through the computational resources and staff contributions provided for the Quest high performance computing facility at Northwestern University which is jointly supported by the Office of the Provost, the Office for Research, and Northwestern University Information Technology.

\appendix
\section{Convergence of discretized equations of motion} \label{convergence}
\begin{figure*}[htb]
    \centering
    \begin{tikzpicture}
        \node at (0,6.4) {\includegraphics[width=0.45\textwidth]{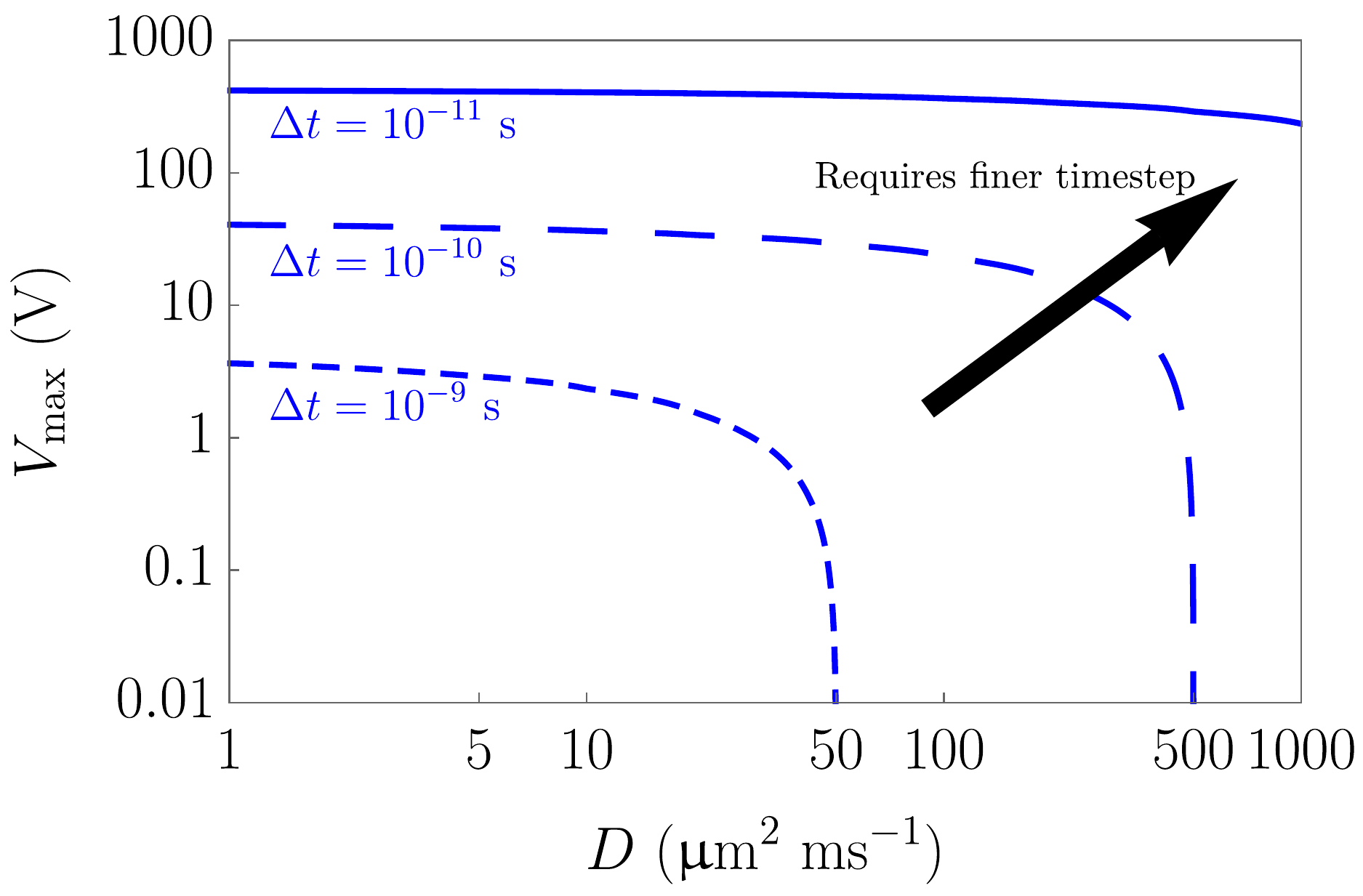}};
        \node at (8.5,6.4) {\includegraphics[width=0.45\textwidth]{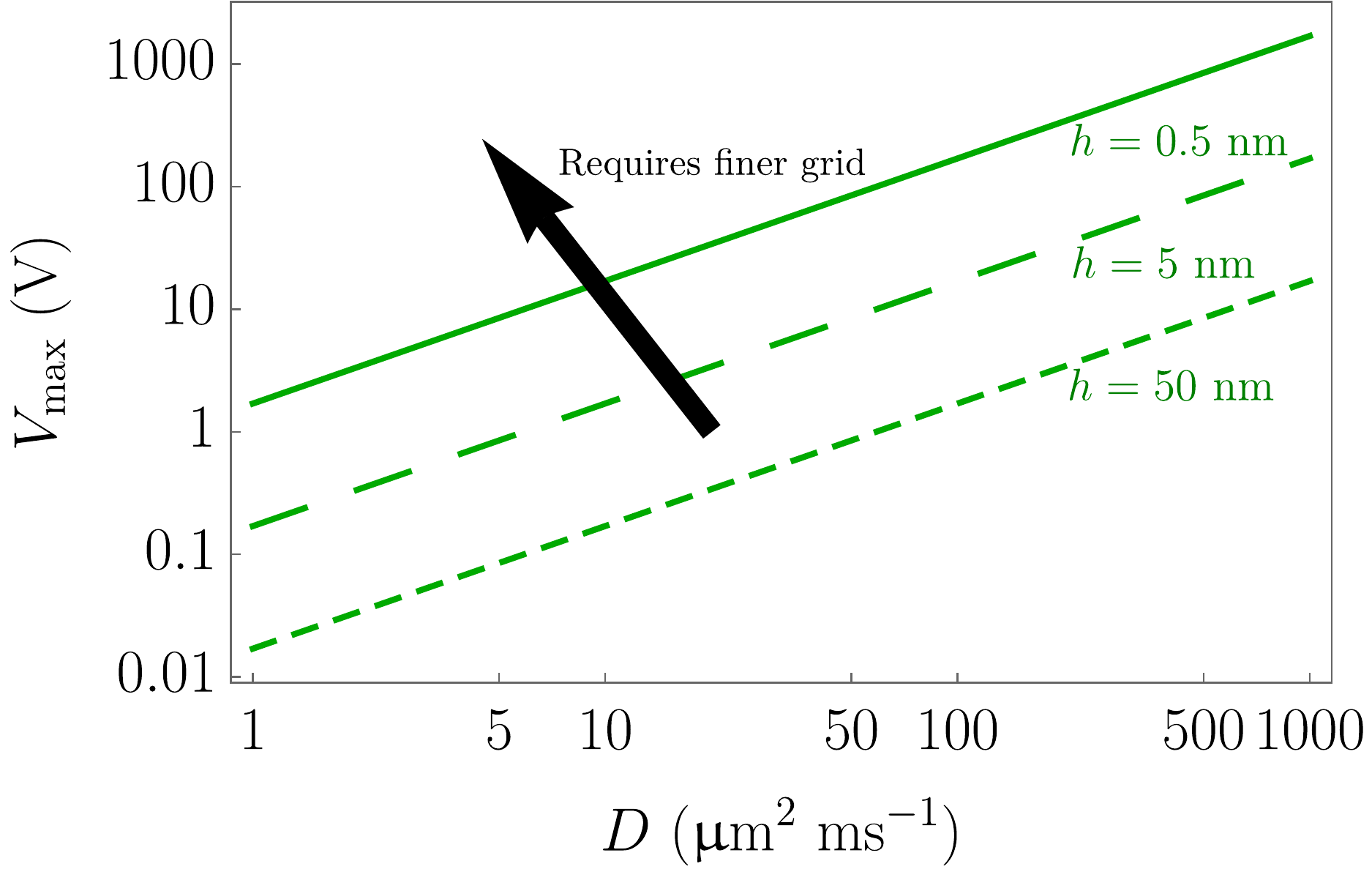}};
        \node at (-3.8,8.6) {(a)};
        \node at (4.7,8.6) {(b)};
    \end{tikzpicture}
    \caption{Stability of temporal and spatial discretization.
(a) Blue curves and (b) green curves represent upper bounds for both the driving amplitude \(V_\text{max}\) and diffusion constant \(D\) needed for well-converged discrete-time and discrete-space currents, respectively.
    All the data points shown were generated with \(U\) as the driving protocol.
    We assume that time steps \(\Delta t\) satisfying Eq.~\eqref{temporalbound} and grid spacings \(h = h^*\) are the marginal values separating reliable results from those which may become unstable.
    The voltage and diffusion constant which results in these marginal discretizations are shown for three choices of \(h\) and \(\Delta t\).}
    \label{stlimits}
\end{figure*}

Both continuous-space and discrete-space methods are approximations that converge to continuum Fokker-Planck dynamics as the discretization size (\(\Delta t\) or \(\Delta h\)) decreases.
We compare the relative merits of these methods here by analyzing how the necessary discretization depends on the diffusion constant and the strength of the external driving.
In short, there can be computational advantages to either method depending on the particular values of \(V_{\rm max}\) and \(D\).

When discretizing time, an upper bound on the acceptable time step \(\Delta t\) may be estimated by requiring that the particle displacement over that time step not grow too large.
We identify the terms in Eq.~\eqref{discretizedLangevin} as a deterministic displacement \(\mathbf{x}_\text{det} = \mu\Delta t\,\mathbf{f}(\mathbf{x}_i)\) and a stochastic displacement \(\mathbf{x}_\text{stoc} = \sqrt{2D\Delta t}\,\boldsymbol{\eta}_i\) and require as a rough heuristic that both deterministic and stochastic displacements are no more than one-hundredth the size of the simulation box.
That is to say \(\Delta t\) must be sufficiently small to ensure
\begin{equation}
    \max\left(\frac{x_\text{det}}{x_\text{max}}, \frac{x_\text{stoc}}{x_\text{max}}, \frac{z_\text{det}}{z_\text{max}}, \frac{z_\text{stoc}}{z_\text{max}}\right) \leq 0.01. \label{temporalbound}
\end{equation}
For a fixed time step \(\Delta t\), the heuristic constraint of Eq.~\eqref{temporalbound} will only be satisfied if the diffusion constant \(D\) and the maximum driving amplitude \(V_{\rm max}\) \((\equiv \max_t |T(t)|)\) are not too large.
For system thickness \(z_{\rm max} = 1\)~\si{\um}, driving amplitudes up to 10~\si{V}, and diffusion constants of 1-100~\si{\um\squared\per\ms}, which are reasonable values for experimental systems involving electron ratchets~\cite{kedem2017light}, we find that a modest time step on the order of 100~\si{ps} is sufficient for accurate simulation.
Accessing larger diffusion constants or larger \(V_{\rm max}\) requires smaller time steps, as illustrated in Fig.~\ref{stlimits}.

Similar to the temporal discretization, the grid spacing for the lattice model must be sufficiently small to ensure convergence.
Although precise limits on the acceptable grid spacing depend heavily on the energy landscape, the minimum requirement is that the hopping rates between neighboring sites of Eq.~\eqref{eq:rates} cannot be negative.
This constraint sets an upper bound for the discretization of the lattice with the largest allowable grid spacing \(h^*\) being the one which causes the smallest rate to drop to zero:
\begin{equation}
    h^* = \min_i\frac{2D}{\mu|f_i|}.
\label{eq:criterion}
\end{equation}
Any \(h>h^*\) involves unphysical negative rates, thus severely affecting the accuracy of the discretization.
If the driving force \(V_\text{max}\) is too large, however, \(h^*\) becomes so small that converged calculations require too fine a grid to be computationally competitive with the Langevin approach.
As shown in Fig.~\ref{stlimits}, the discrete-space computations become more favorable---converging with a coarser grid---as the diffusion constant increases, while the discrete-time simulations show the opposite trend, requiring a smaller time step and hence a more expensive calculation.
Roughly speaking, discrete-time simulations are preferable given a driving voltage beyond 1~\si{V}, whereas the discrete-space calculations are more attractive underneath that threshold.

\begin{figure*}[htb]
    \centering
    \begin{tikzpicture}
        \node at (0,6.4) {\includegraphics[width=0.45\textwidth]{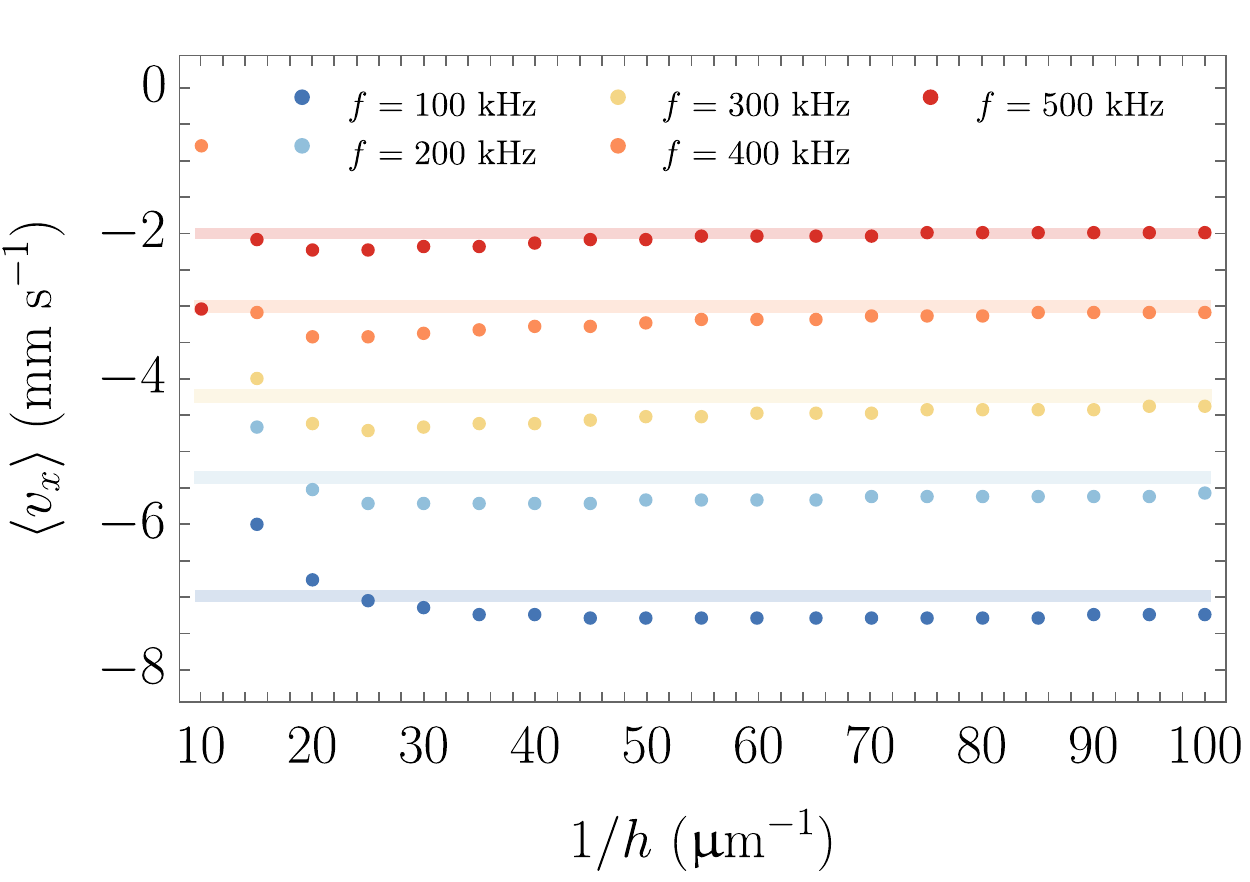}};
        \node at (8.5,6.4) {\includegraphics[width=0.45\textwidth]{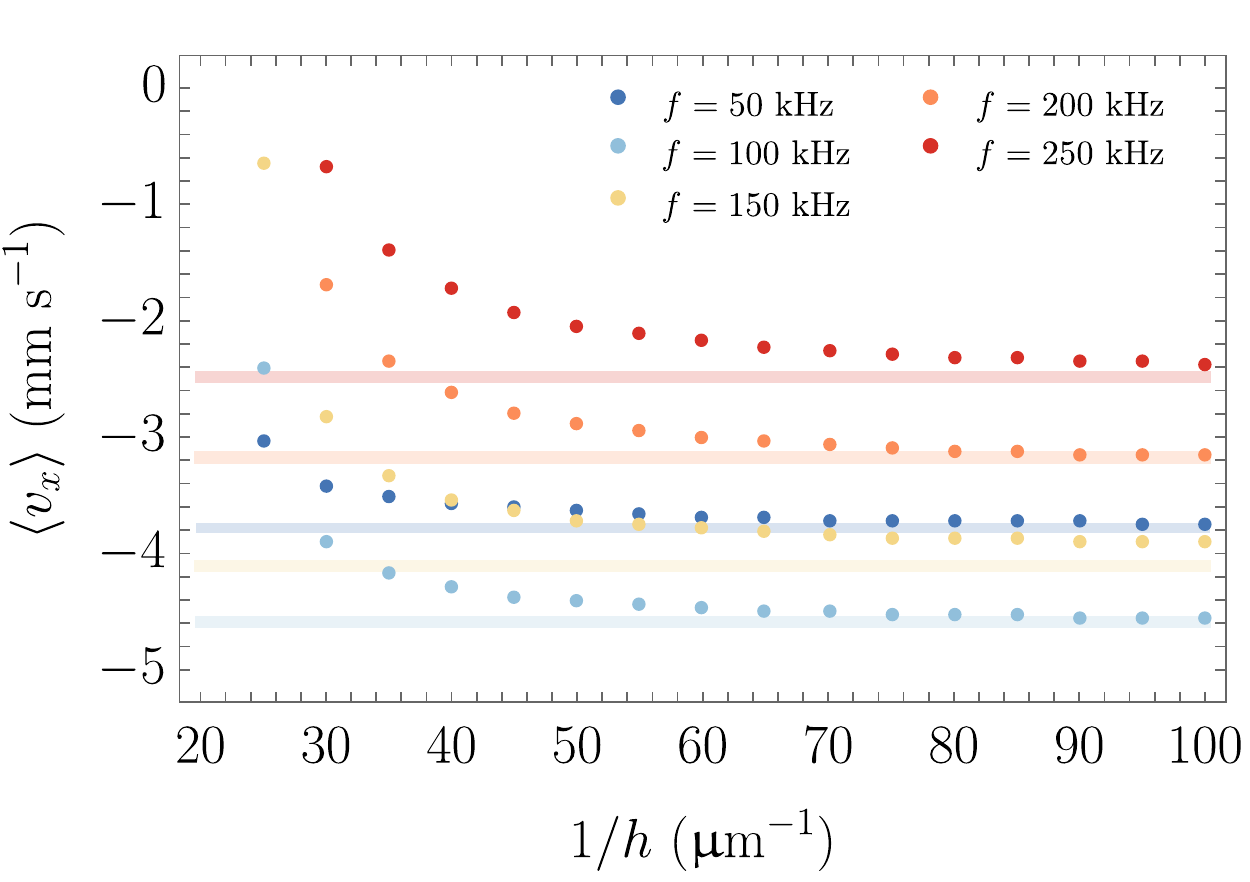}};
        \node at (-3.8,8.6) {(a)};
        \node at (4.7,8.6) {(b)};
    \end{tikzpicture}
    \caption{Convergence of spatial and temporal discretization.
Average horizontal particle velocities computed from spectral calculations are plotted with dots for potential (a) \(U\) and (b) \(U_2\).
As the grid spacing shrinks, currents approach those computed from an average of 512 independent 10~\si{\ms} Langevin trajectories with a time step \(\Delta t\) of 35~\si{\ps}.
Each shaded rectangle shows one standard error around the corresponding Langevin currents.
}
    \label{conv}
\end{figure*}

For various driving frequencies and both driving protocols \(U\) and \(U_2\), Fig.~\ref{conv} demonstrates the convergence of discrete-space currents toward the continuum limit as the grid spacing \(h\) is decreased.
The value of \(h\) chosen in our calculations, \(1/100~\si{\um}\), is more than sufficiently small for the convergence of spatially discretized currents, as conveyed by the figure.

\section{Scaled cumulant-generating function from Markov process} \label{SCGF}
We derive Eq.~\eqref{eq:perron} by a limiting procedure.
Let the time-periodic rate matrix \(\mathsf{W}\) have a period \(\tau\) consisting of \(\nu\) equal segments of time-constant rate matrices \(\mathsf{W}_1, \mathsf{W}_2, \dotsc, \mathsf{W}_\nu\). 
The SCGF \(\psi_X(\lambda)\) of a Markov jump process with finite state space \(\Sigma = \{1, \dotsc, |\Sigma|\}\) for the random variable 
\begin{equation}
X = \frac{1}{T}\sum_{i,j\in\Sigma}d_{ji}q_{ji}(0,T)
\end{equation}
is given by
\begin{equation}
\psi_X(\lambda) = \frac{1}{T}\ln\max{\rm eig}\prod_{\kappa=0}^{\nu-1} e^{(T/\nu)\mathsf{W}_{\nu-\kappa}(\lambda)},
\end{equation}
where the tilted rate matrices \(\mathsf{W}_\kappa(\lambda)\) satisfy 
\begin{equation}
[\mathsf{W}_\kappa(\lambda)]_{ji} := [\mathsf{W}_\kappa]_{ji}e^{\lambda d_{ji}}.
\end{equation}

\textit{Proof.} Discretize the jump process by considering snapshots of the system at intervals \(\Delta t = T/N\), \(N \to \infty\), as a Markov chain. Similarly discretize \(X\), with
\begin{equation}
TX = \lim_{N\to\infty}\sum_{k=2}^Nd_{\sigma_{k-1},\sigma_k},
\end{equation}
where \(\sigma_k\) denotes the state of the system at time \(t = k\Delta t\). Hence, by definition,
\begin{align}
    \langle e^{\lambda TX}\rangle &= \lim_{N\to\infty}\sum_{\sigma_1,\dotsc,\sigma_N}[\mathsf{T}_\nu(\lambda)]_{\sigma_N,\sigma_{N-1}}\cdots[\mathsf{T}_1(\lambda)]_{\sigma_2,\sigma_1}p_{\sigma_1}\\
    &= \lim_{N\to\infty}\mathbf{1}^\top\mathsf{T}_\nu(\lambda)^{N/\nu}\cdots\mathsf{T}_1(\lambda)^{N/\nu}\mathbf{p} \nonumber,
\end{align}
where \(p_i\) is the probability of starting out in state \(i\), \(\mathbf{p}\) is the vector with components \(p_i\), \(\mathsf{T}_\kappa(\lambda)\) the tilted transition matrix with elements satisfying
\begin{equation}
[\mathsf{T}_\kappa(\lambda)]_{ji} := [\mathsf{T}_\kappa]_{ji}e^{\lambda d_{ji}},
\end{equation}
where \([\mathsf{T}_\kappa]_{ji}\) is the transition probability from the state \(i\) to the state \(j\) subject to the rate matrix \(\mathsf{W}_\kappa\), and \(\mathbf{1}\) the vector with elements all ones. In the large-\(N\) limit, by the Perron-Frobenius theorem, 
\begin{equation}
\langle e^{\lambda TX}\rangle = \lim_{N\to\infty} c_\lambda\max{\rm eig}\prod_{\kappa=0}^{\nu-1}\mathsf{T}_{\nu-\kappa}(\lambda)^{N/\nu},
\end{equation}
for some insignificant factor \(c_\lambda\), which gives
\begin{align}
    \psi_X(\lambda) &=  \frac{1}{T}\ln\max{\rm eig}\prod_{\kappa=0}^{\nu-1}\mathsf{T}_{\nu-\kappa}(\lambda)^{N/\nu}.
\end{align}
We finally identify the elements of the tilted rate matrices \(\textsf{W}_\kappa(\lambda)\). The \(\textsf{W}_\kappa(\lambda)\) are defined as generators of the tilted transition matrices \(\textsf{T}_\kappa(\lambda)\), and their elements may hence be found by a Taylor expansion in \(\Delta t\), with
\begin{align}
    [\mathsf{T}_\kappa(\lambda)]_{ji} &= [\mathsf{T}_\kappa]_{ji}e^{\lambda d_{ji}} \approx \delta_{ji} +  [\mathsf{W}_\kappa]_{ji}e^{\lambda d_{ji}}\Delta t\\
    &=: [\mathsf{I} + \mathsf{W}_\kappa(\lambda)\Delta t]_{ji} \nonumber,
\end{align}
and inspecting the coefficient of \(\Delta t\) in the final equality yields, as desired,
\begin{equation}
[\mathsf{W}_\kappa(\lambda)]_{ji} := [\mathsf{W}_\kappa]_{ji}e^{\lambda d_{ji}}.
\end{equation}

\bibliography{biblio.bib}

\end{document}